\documentclass[
reprint,  
dcolumn, 
superscriptaddress,
 amsmath,amssymb,
 aps,
prd,
]{revtex4-2}

\usepackage{graphicx}
\usepackage{dcolumn}
\usepackage{bm}

\usepackage{appendix}

\begin{document}

\title{A novel experimental approach to uncover the nature of cosmic-ray Deuterium}

\author{F. Dimiccoli}
\email{francesco.dimiccoli@unitn.it}
\affiliation{University of Trento, Department of Physics, V. Sommarive 14, I-38123, Trento, Italy}
\affiliation{INFN-TIFPA, V. Sommarive 14, I-38123 Povo (Trento), Italy}

\author{F. M. Follega}
\email{francesco.follega@unitn.it}
\affiliation{University of Trento, Department of Physics, V. Sommarive 14, I-38123, Trento, Italy}
\affiliation{INFN-TIFPA, V. Sommarive 14, I-38123 Povo (Trento), Italy}

\date{\today}

\begin{abstract}
Studying the isotopic composition of cosmic-rays (CRs) provides crucial insights into the galactic environment and helps improve existing propagation models. Special attention is given to the secondary-to-primary ratios of light isotopic components in CRs, as these measurements can offer complementary data compared to traditional secondary-to-primary ratios like B/C. Recently, a precision measurement of the Deuterium (D) abundance in CR in the 2-21 GV rigidity range provided by the AMS02 experiment unexpectedly detected an excess of D with respect to its expected secondary nature, opening the field for new measurements at high rigidity to determine how the spectrum evolves and whether there is confirmation of a primary or primary plus secondary origin. While there are theoretical models that attempt to explain this excess, the experimental uncertainties on D production cross-sections and on CR propagation models remain significant, and only new and precise measurements can dissipate existing doubts. In this work we review the current experimental scenario and we propose a dedicated experiment able to extend the D abundance measurement up to 100 GeV/nucl without the need of a magnetic spectrometer, using a multiple scattering based technique for the measurement of particle momentum. The expected performances of the proposed detector were assessed through a dedicated simulation using the GEANT4 package, and its role in the current particle physics scenario is discussed.

\end{abstract}

\maketitle


\section{\label{sec:level1}Introduction}

Studying the flux of cosmic-ray Deuterium (D) is pivotal to probe the mechanisms of cosmic-ray propagation and improve the understanding of the underlying astrophysical processes. For long time, Deuterium was believed to be a pure secondary \cite{coste}, overwhelmingly produced through the spallation of heavier nuclei, mainly $\mathrm{^4He}$ interacting with the interstellar medium (ISM) \cite{PhysRevC.98.034611}.

The understanding of the propagation mechanisms of this element is crucial for multiple reasons.
First, being D significantly lighter with respect to other secondary elements like B, it can effectively probe CR propagation at larger galactic distances \cite{mosk} and measurements of the ratio of D against primaries like p and $\mathrm{^4He}$, could provide insights into the spallation processes and the composition of the ISM. Second, as it is produced by a light primary like $^4$He, it offers complementary insights compared to heavier secondary-to-primary ratios such as B/C, which have traditionally been used as tools to study CR propagation. Eventual discrepancies in diffusion behaviors at different Z may reinvigorate discussions about scenarios which violate the CR propagation universality postulate \cite{Tomassetti_2017}. 
Third, the development of reliable models of CR propagation faces important theoretical challenges, due to the limited experimental knowledge of inhelastic cross section for all the involved nuclei \cite{PhysRevD.96.103005}. Therefore, a measurement of deuteron flux at high energies could be used to constrain these uncertainties.
\begin{figure*}[t]
    \centering
    \includegraphics[width=1.05\linewidth]{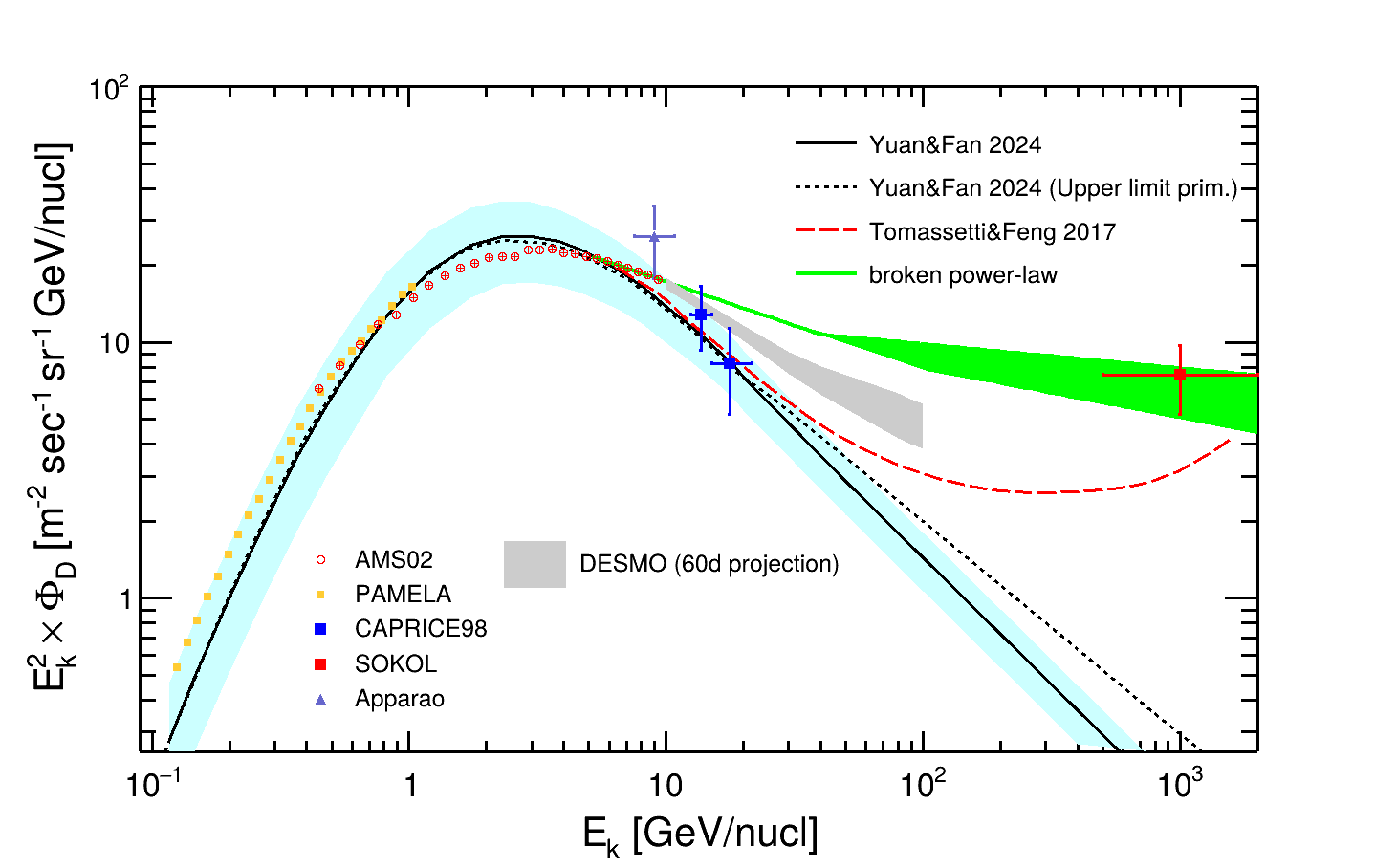}
    \caption{Measurements and models of Deuterium flux as a function of the kinetic energy per nucleon ($E_k$), and multiplied by $E_k^2$, are shown. Data points represent experimental measurement by AMS02 \cite{AMS2024}, PAMELA \cite{Pamela}, CAPRICE98 \cite{caprice},  SOKOL \cite{sokol} and Apparao et al. \cite{apprao1983}. The black lines represent the models obtained by \cite{Yuan_2024}, in the case of purely secondary deuterons (continuous line) and considering the calculated upper limit of primary contribution (dotted). These models were extrapolated at high energies with a single power-law trend above 20 GeV/nucl. The red dashed line shows the model presented in \cite{Tomassetti_2017}. The green line + band represents the extrapolation of AMS02 data calculated as described in Section \ref{scrutinize}. The gray band represents the projection of an hypothetical measurement by the DESMO detector (Section \ref{concept}) in 60 days, based on the performance estimated in the following section of this work, assuming a deuteron flux dependence $\propto E_{k}^{-2.7}$ (see Section \ref{perform}). }
    \label{fig:d_over_p_flux}
\end{figure*}
 Moreover, improving the understanding on the propagation process of CR in the galaxy, and thus of the property of the galactic environment itself, has also profound implications regarding the prediction of the amount of light anti-nuclei (e.g. $\bar{p}$, $\bar{D}$) produced by spallation reactions. This estimation is pivotal in the indirect searches for Dark Matter (DM), since the exotic production of light anti-nuclei in the DM annihilation process is one of the most promising channels for its indirect detection. In particular, the production of $\bar{D}$ in CR spallation reaction is kinematically disfavored. As such, theoretical predictions show that the $\bar{D}$ is an almost background-free channel for indirect DM searches \cite{antid_1}. In this context, the study of the Deuterium propagation assumes also the role of a proxy for the behavior of the $\bar{D}$ nuclei in the galaxy.

All the arguments above assume that deuterons predominantly originate as a secondary component. However, the unambiguous identification of deuterons as pure secondary cosmic-rays requires measuring the rigidity dependence of its flux at high energies and in particular the ratio with its supposed progenitor, the D/$^4$He ratio. Until recently, this has not been achieved due to the experimental difficulties involved.

Recent measurements from the Alpha Magnetic Spectrometer (AMS02) on the International Space Station have provided detailed spectra of cosmic-ray deuteron up to 21 GV \cite{AMS2024}. Surprisingly, the deuteron spectrum at high energies is significantly harder than what is typically expected for secondary species. The rigidity dependence of these particles closely resembles that of protons (p), yet it differs significantly from that the of $\mathrm{^3He}$, another isotope produced by $\mathrm{^4He}$ spallation. This feature confirms the presence of an excess of high energy D with respect to the one expected from secondary production alone, suggesting the existence of a primary-like deuteron component. Given the implications of these new data on the phenomenology of CR propagation, the situation deserves more clarification on the experimental side, possibly extending the charted rigidity range for this measurement.

In this work, we review the current state of the measurements and the models concerning cosmic-ray deuterons, emphasizing on the hints of a primary-like nature. In the second part, we propose DESMO (DEuterio con Scattering MultiplO), a ballon-borne experiment, designed to measure deuteron flux at higher energies.

\section{Scrutinizing measurements and models about high-energy deuteron flux}\label{scrutinize}

The current experimental and theoretical knowledge about high-energy D cosmic-ray flux is summarized in Figure \ref{fig:d_over_p_flux}. As can be seen,  AMS02 is the only high-precision measurement available at energies that exceed the domain of solar modulation, probing the spectral shape of this nucleus above 1 GeV/nucl. As introduced above, the hardening of the deuteron spectrum toward the higher energies reported by AMS02 was unexpected, because traditional astrophysical models do not predict significant D production outside of the Big Bang nucleosynthesis (BBN) \cite{bbnuc}. Recent theoretical studies \cite{Yuan_2024} made use of the  numerical code GALPROP \cite{galprop} to simulate the D galactic diffusion and showed that, considering all the contributions from fragmenting nuclei up to Z=28, it is possible to explain the AMS02 measurement within a secondary production framework. The model is compatible with AMS02 results within errors, dominated by the uncertainty on diffusion parameters and on nuclear cross sections. This result underlines the importance of the contribution given by the fragmentation of heavy nuclei to the Deuterium flux. However, considering the scarcity of direct cross-section measurements, this interpretation of the measured excess of deuterons is currently object of debate across the scientific community \cite{arx}. Nevertheless, authors from \cite{arx} obtains an upper limit for an hypothetical primary component (D/H $\le$ 1.6 $\times$ 10$^{-3}$) that could explain the AMS02 measured data, this would imply a much higher abundance of Deuterium  with respect to the one expected from BBN and/or a much more efficient acceleration mechanism for D. In Figure \ref{fig:d_over_p_flux}, we reported also these predictions, extrapolated above 20 GeV/nucl, assuming a constant power-law dependence.
Currently, other measurements of the deuteron flux at high energies exist, independently conducted over the years by various experimental apparatus. We firstly report the result of the deuteron flux obtained by Apparao et al. \cite{apprao1983} and reported in other works (\cite{sokol},\cite{Tomassetti_2017}), which is compatible with the AMS02 measurement. At higher energies, only few experimental points with large uncertainties are available. The measurements provided by CAPRICE98, shown in Figure \ref{fig:d_over_p_flux}, cover an higher energy range and are in better agreement with the extrapolated model prediction with respect to the trend of the AMS02 data. However, despite the large error bars, they still hints toward an excess of high-energy deuterons. 
Finally, indications of anomalous D abundance at extremely high-energy were provided by the SOKOL satellite experiment \cite{sokol}. Discriminating between the different shapes of the hadronic cascades induced by D and p nuclei at very high energies, they were able to measure a fraction of D in cosmic hydrogen of (37.0 $\pm$ 7.6)\% between 1 and 4 TeV of total energy. We converted this measurement in an absolute D flux measurement using as an anchor the integral of the proton flux measured by the DAMPE experiment \cite{DampeP} in the same energy range. The result of this calculation, converted in kinetic energy per nucleon, is also shown in the Figure. Moreover, authors of \cite{Tomassetti_2017} have shown that such measurement can be explained in terms of standard mechanisms of D production (red long-dashed line in Figure \ref{fig:d_over_p_flux}) assuming a different acceleration mechanisms for light and heavy primaries. If confirmed, this scenario would have profound implications, since the secondary to primary ratios involving the heavy primaries would not be able to place constraints on the production of light isotopes or anti-particles. 
On this topic, we propose an extrapolation of the AMS02 data, shown in green in Figure \ref{fig:d_over_p_flux}, that interestingly enough aligns with the SOKOL measurement. We obtained their spectral index ($\gamma = -2.34 \pm 0.01$) from a single power-law fit above 5 GeV/nucl, and introduced a high-energy spectral index change, $\Delta \gamma$, ranging from 0.15 to 0.20, as observed by AMS02 in primary nuclei \cite{PhysRevLett.119.251101}. This spectral shape was modeled using a broken power-law with a rigidity break point between 80 and 200 GV. The green band represents the variability of the model under these parameter variation and aligns with the SOKOL observations.

Basing on the current theoretical and experimental knowledge of the topic, we depict three scenarios: a) D is a secondary species or it has a limited primary-like component, as shown in \cite{Yuan_2024}; b) D is a secondary species, but produced by primaries accelerated with different mechanisms, as hypothesized in \cite{Tomassetti_2017}; c) D follows a hard primary-like power-law spectral shape that continues the trend measured by AMS02. Only through a precise measurement of the D flux spectral index at higher energies the distinction between these three scenarios is possible. 

To address this challenge, we propose a new experimental strategy that builds on recent advancements in detection technology, keeping as a cornerstone stringent constraints on mission complexity and cost. Leveraging on the methodology shown in \cite{DimiccoliFollega2024}, which utilizes multiple scattering for isotopic identification, we propose DESMO for a two-months balloon flight.
DESMO is a new and efficient solution for the study of cosmic-ray isotopes at high energies with a relatively compact and light design, operating in the rigidity range between 20 and 200 GV. With the gray band in Figure \ref{fig:d_over_p_flux} we show a projection of a 60 days long DESMO measurement based on the performances estimated in the following sections and its ability to discriminate between the aforementioned (a), (b) and (c) scenarios, finally shading light on the origin of cosmic-ray deuterons.

\section{The DESMO Concept}\label{concept}

The concept behind DESMO is to obtain a lightweight, relatively cheap and robust detector able to provide a measurement of the D flux, that can operate as light payload in a long duration balloon flight. To fulfill this vision, we imposed stringent requirements on the total detector mass (below 100 kg), on the total tracking area (below 2 m$^2$) and on the total detector length (below 1.6 m), which makes DESMO at least an order of magnitude smaller than future and current cutting-edge experiments like AMS02.

\begin{figure*}[t]
\centering
\includegraphics[width=0.8\textwidth]{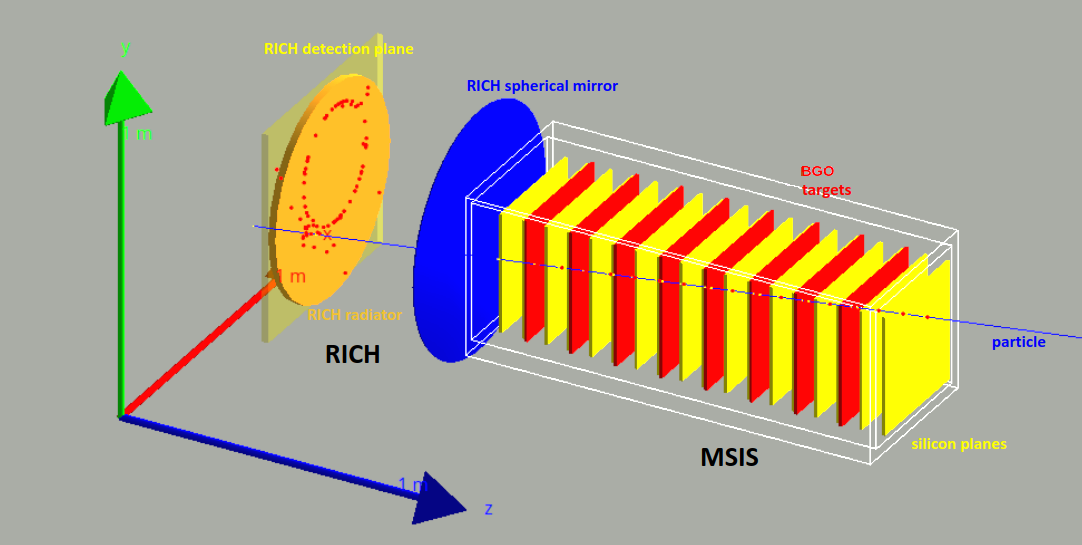} 
\caption{Rendering 3D of the DESMO detector, with its principal components labeled: RICH and MSIS sub-detectors (see the text).
A simulated D event of generated energy 50 GeV/nucl is also shown. The red points represent the hits of the primary particle and of the Cherenkov photons produced by the particle passage, measured by the detector active areas. The yellow points represent the multiple scattering interactions. In particular, the reflected Cherenkov ring is visible on the RICH detection plane. The tracks of secondary particles were not drawn for visualization purposes. The green, red and blue arrows show a 1 m length in the x,y and z directions as a visual scale reference.}
\label{fig:desmopic}
\end{figure*}

 This can be achieved using a novel technique for isotopic identification in cosmic-rays, based on multiple scattering, which exploits the different angles at which isotopes of the same velocity but different masses are scattered. This approach has already been proposed for measurement of particle momentum \cite{Agafonova_2012}, and offers an interesting alternative to traditional magnetic spectrometers, which face limitations in resolution and complexity when targeting high energies. As shown in \cite{DimiccoliFollega2024}, it is possible to exploit the dependence of the multiple scattering effect with the particle momentum to obtain a measurement of the latter without the need for strong magnets to bend particle trajectories. This, in conjunction with a measurement of particle velocity, allows for a mass measurement and thus for an identification of isotopes. 
DESMO is composed of two fundamental modules: the first one measures the velocity of the incoming particle, the second one measures the average displacement induced by the multiple scattering on the particle trajectory. Its final design is shown in Figure \ref{fig:desmopic}. The first module, a particularly compact design of Ring Imaging Cherenkov detector (RICH), makes use of a spherical mirror and fulfills the first task, besides estimating the incoming particle charge Z. 
The second module, the Multiple Scattering Isotope Separator (MSIS), is able to separate different masses among nuclei of same Z and velocity, providing also an independent charge measurement.

\subsection{The MSIS detector}\label{MSIS}

The multiple scattering approach involves tracking the deflections of charged particles as they pass through a series of thin targets contained in several measuring stations. Every measuring station consists of two tracking planes and of a high-density and high-Z target, to induce the deviation of the particle trajectory by multiple scattering. We call this fundamental unit of the detector the PPT module (Plane-Plane-Target). Each PPT module measures the incoming particle trajectory and deviates it for the next measurement. The MSIS detector is composed of eight identical PPT modules, and of two additional tracking planes placed at its bottom to measure the deviation induced by the last station and obtain a total of eight measurements. In this way, a tower of N different stations can provide the measurement of N different trajectories and thus N-1 scattering angles $\theta_i$.
This method has been validated through extensive simulations using the GEANT4 package \cite{geant4}, demonstrating its capability to identify Z=1 isotopes at high energies \cite{DimiccoliFollega2024}.  
 The angle $\theta_i$ is measured using the linear displacement d$_i=S\tan{\theta_i}\approx S\theta_i$ of the hit measured by the first tracking plane with respect to the linear extrapolation of the trajectory measured by the precedent PPT module, where S=6 cm is the spacing between two subsequent PPT modules. The eight displacement measurements are combined to obtain an average displacement $d$, which is proportional to the multiple scattering angle $\theta_{MS}$:

\[
d = \frac{\sum_i{d_i}}{8} \approx \frac{\sum_i{S \theta_i}}{8} \approx S \theta_{MS}
\]

 The silicon planes of each PPT modules are all equal, each measuring 33 cm $\times$ 33 cm and 3 mm thick, with a 6 cm separation between them. Typical values of displacements for (10-100) GeV/nucl p and D in MSIS are respectively in the range between 10 $\mu$m and 200 $\mu$m, so high spatial resolution  is critical for accurately determining scattering angles. This can be achieved with a matrix by ALPIDE pixel sensors (28$\times$28 $\mu$m$^2$), which, determining the centroid of the pixel cluster, can reach a sub-pixel spatial resolution of the order of 5-10 $\mu$m \cite{AGLIERIRINELLA2017583}, with low material budget per tracking plane. Tracking detectors of this kind have already been produced and integrated in space detectors \cite{ester,savino}.
 A target material, with high density and high-Z properties, is essential for inducing significant scattering, which is then measured by the tracking planes. An active material like a crystalline scintillator, such as Bismuth-Germanate (BGO) with an effective $Z$ of 74 and a density of 7.13 g/cm$^3$ \cite{bgo}, allows for energy deposition measurements and provides a fast signal that can serve as a trigger for DESMO.
For these reasons, a 14 mm thick BGO target is positioned next to the second tracking plane of every PPT module. 
By measuring energy deposition, each target provides an independent measurement of the incoming particle's charge $Z$.
The majority of nuclei in the DESMO target energy range are in the Minimum Ionizing Particle regime and penetrate many of such targets before having a significant chance for inelastic interactions. Thus, the combinations of the energy deposition measurements from the first two BGO targets can be used to obtain a precise charge measurement. From studies performed on the GEANT4 simulation of the MSIS we estimated a charge resolution of $\Delta Z/Z \sim 0.04$, lowering the contamination from Z$\geq$2 nuclei on the D signal to negligible levels. Additionally, BGO offers the opportunity for an optional target segmentation, enabling a rough trajectory estimate to trigger only nearby silicon pixel modules, significantly reducing their readout power requirements.

The MSIS design specifications were determined through an optimization process detailed in Appendix \ref{sec:optim}, aimed at maximizing the precision of the average displacement $d$, while adhering to the design requirements outlined at the beginning of this section.
The performance of this design has been validated through simulations, demonstrating its capability to effectively separate protons and deuterons at high energies. More details on this can be found in \cite{DimiccoliFollega2024}. The optimization of the PPT module design involved extensive simulations using the GEANT4 package and additional tools reproducing the digital signal of the pixels \cite{MAGER2016434}. These simulations accounted for various factors, including the energy range of interest (10-100 GeV/nucl), the spatial resolution of the silicon planes, and the scattering properties of the target. A detail study is shown in Appendix \ref{sec:optim}. These studies showed that the DESMO detector could achieve a significant separation power (defined as in \cite{Vuchkov2001}) for Z = 1 isotopes, making it a promising tool for a cosmic-ray D measurement.

\subsection{The RICH detector}\label{RICHsection}

Particle kinetic energy per nucleon $E_{k}$, measured in GeV/nucl, can be obtained from a measurement of particle velocity ($\beta$ = v/c), through the relation

$$E_{k} = \frac{E_{kin}}{N_{nucl}} = m_{nucl} (\gamma -1) $$
where $m_{nucl}$ is defined as
$$m_{nucl} = \frac{N_p m_p + N_n m_n}{N_p + N_n} $$
where $N_{nucl}$ is the number of nucleons, $m_{nucl}$ is the mass of the average nucleon,  with $N_p$ ($N_n$) and $m_p$ ($m_n$) being respectively the number and mass of the protons (neutrons).
A precise velocity measurement is needed in the 10-100 GeV/nucl energy range. The combined requirements of precision and compactness in the design can be achieved only using a Cherenkov detector \cite{viehhauser_weidberg_2024}.

A compact design Ring Imaging CHerenkov (RICH) detector was devised for this purpose, based on a 2 cm thick layer of Silica Aerogel (SiO$_2$) radiator with spectral index of n=1.05 and density of 0.2 g/cm$^3$ \cite{PWang_1994,Giovacchini:2023ixx,PEREIRA}.

\begin{figure}[t]
\centering
\includegraphics[width=0.44\textwidth]{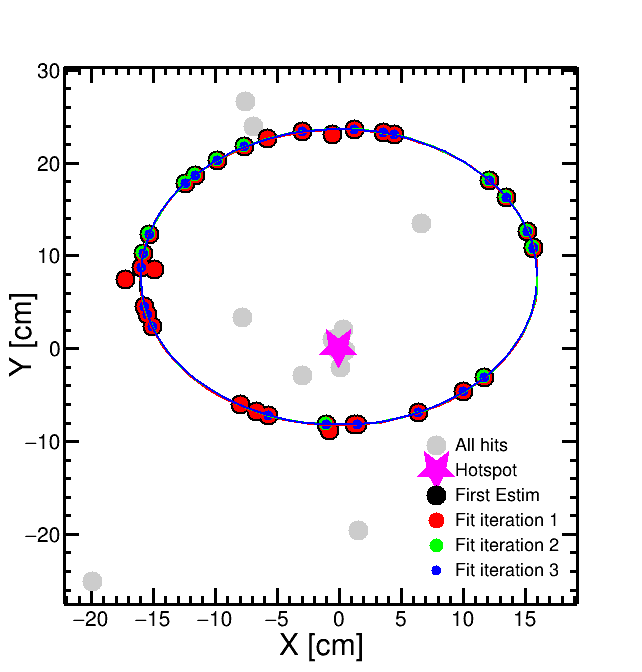}
\includegraphics[width=0.46\textwidth]{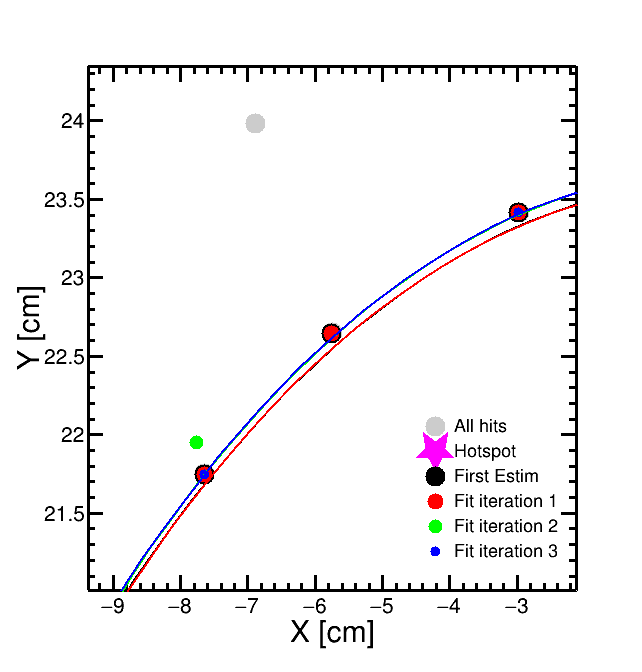}
\caption{Iterative elliptical fit on a typical distribution of detected Cherenkov photons hits (left). A zoomed view (right) of the same event is also shown. At every iteration, a different set of hits is pooled and the fit is adjusted. In this plot, hits generated by the dark count rates of the SiPM are not shown for graphical purpose (see more details in Appendix \ref{sec:sipm}).}
\label{fig:richfit}
\end{figure}
 A spherical mirror (featuring 1 m curvature radius and 50 cm diameter, placed at a distance d$_{M}$ = 50 cm below the radiator) refocuses back Cherenkov photons onto a position sensing SiPM  based detector array (50x50 chips, 10×10 mm$^2$ each) positioned right below the radiator. Each chip  (Linearly Graded SiPM - LG-SiPM \cite{lgsipm}) can obtain position measurement through center-of-mass estimation of the electron avalanche triggered by the incoming photon, reaching spatial resolution of the order of 50 $\mu$m. A more detailed discussion of the spatial resolution attainable by this chip and other commercial and future alternatives is reported in Appendix \ref{sec:sipm}. 
Figure \ref{fig:desmopic} shows graphically the proposed design of RICH integrated in the DESMO detector. The compact design of RICH aligns well with the requirements of a typical balloon-borne experiment. Such experiments generally include a thermal control system to maintain a stable operating temperature during both the ascent and flight phases \cite{helixprc,gaps}. For DESMO, this system is essential—not only to dissipate heat from the MSIS readout electronics but also to ensure the mantainment of the structural properties of the RICH mirror. Additionally, the operating temperature directly influences the performance of the RICH readout plane (see Appendix \ref{sec:sipm} for details). A thermal control system capable of maintaining a constant operating temperature of 0$^\circ$ is the reference choice for DESMO. 
The detector's concept was initially tested for perpendicular tracks \cite{DimiccoliFollega2024}. This study extends the evaluation to a realistic cosmic-ray scenario simulating single charged high energy particles with inclinations up to 20° (maximum angular acceptance of the MSIS detector below) and generic impact point, using the GEANT4 simulation toolkit.
During their passage single charged particles emit 50-100 photons by Cherenkov radiation resulting in 30-50 hits on the photo-detection plane. Although, this is a quite low number of photons (<1ph/SiPM), thanks to the refined fitting procedure described below, it is possible to correctly extract the particle properties, even in presence of dark noise counts (see Appendix $\ref{RICHsection}$).

In the general case, the Cherenkov cones are refocused by the mirror in an ellipse on the detector plane described by ${X^2/S_A^2 + Y^2/S_B^2 = 1}$, with
\begin{eqnarray*}
    X &=& (x-X_0) \cos\alpha + (y-Y_0)\sin\alpha \,, \\
    Y &=& -(x-X_0) \sin\alpha + (y-Y_0)\cos\alpha \,,
\end{eqnarray*}
where $X_0$, $Y_0$ parameters are the ellipse center coordinates, the semi-axes $S_A$, $S_B$, and the angular tilt $\alpha$ between the ellipse ``horizontal" axis and the horizontal direction. This curve is reconstructed by the RICH readout plane. Cherenkov photon hits are fitted to extract ellipse parameters ($S_A, S_B$, $X_0, Y_0$, $\alpha$). The precision on the parameters is limited by the total number of photons collected, which is primarily influenced by the amount of radiator traversed by the charged particle and by the quantum efficiency of the SiPM \cite{lgsipm,AMBROSI}. For the latter we assumed a constant value of 50\% (see Appendix \ref{sec:sipm}).  Moreover, the occurrence of large scattering phenomena like the Rayleigh scattering or auto-absorbing effect of the radiator can further limit the amount of useful photons. Such effects are accounted for in the simulation, modeling the optical properties of Silica Aerogel material \cite{iwasi}. Also imperfection of the mirror surface were taken into account, adding a smearing in the reflected photon direction of the order of 0.5 mrad \cite{CISBANI2003305}.
\begin{figure}[ht]
\centering
\includegraphics[width=0.49\textwidth]{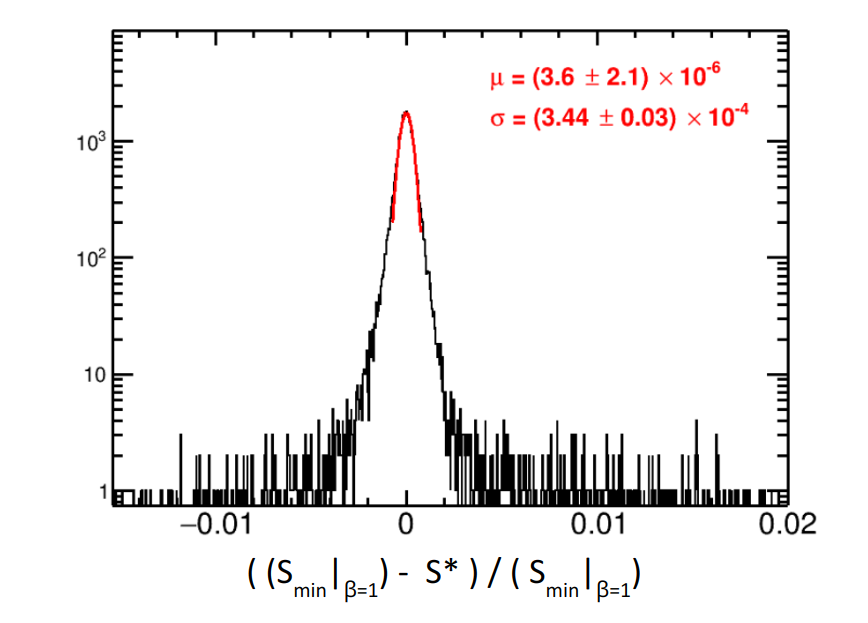}
\caption{The performance of the BDT calibration procedure evaluated as residual distribution of the BDT response (S$^*$) with respect to the regression target variable (minor semi-axis of the measured ellipse S$_{min}|_{\beta=1}$) are shown. They are obtained with a test sample of $\beta \sim$ 1 particles generated with generic inclinations (0$^\circ$-20$^\circ$) and impact points on the radiator. The red curve represents a Gaussian fit of the core of the distribution. }
\label{fig:bdtperf}
\end{figure}
Hits coming from photons not correctly focused, because of scattering phenomena or mirror imperfections, can be excluded from the analysis with an iterative fit procedure.  The procedure can be summarized as follows: 
firstly, a filtering procedure of the recorded hits is performed leveraging on information such as the crossing point of the particle with the RICH readout plane ("Hotspot") and the expected center of the ellipse, extrapolated from the track measured by MSIS.
In particular, hits inside 1 mm radius from the Hotspot are not considered. Moreover, the distribution of distances d of the hits from the ellipse center allows an effective filtering of random dark current hits (see Appendix \ref{sec:sipm}). Only hits with d within 2 cm from the distribution mode are retained. A first fit is performed on the remaining hits. Then, a series of subsequent iterations are performed repeating the fit, but considering only the hits in the vicinity of the previous estimation of the ellipse. This selection is performed by studying the distribution of the distances from the available hits and the fitted ellipse. The standard deviation $\sigma$ of the resulting histogram is calculated and only hits within 2$\sigma$ from zero are retained. It is worth noticing that previously discarded hits can be recovered if their distance is below threshold in the subsequent iterations. The procedure stops whenever for two subsequent iteration the same collection of hits is selected, usually requiring 3-4 iterations. A minimum of 15 hits per fit is required to provide an accurate estimation of the ellipse semi-axes reconstruction (i.e. $10$ $\mu$m) and only fits with a reduced $\chi^2$ value lower than 3 are accepted in the analysis. These parameters were optimized on the trade-off between reconstruction accuracy and efficiency loss, the latter ranging from 30\% to 45\% with this final choice of parameters.\\
Figure \ref{fig:richfit} shows the graphical rendering of the Cherenkov photon hits created by an high-energy (80 GeV/nucl) deuteron crossing the RICH detector and the elliptic fit performed on them.
The minor semiaxis S$_{min}$=min\{S$_A$,S$_B$\} depends both from particle $\beta$ and particle trajectory. It can be written thus as S$_{min}$ ($\theta$,$\phi$,\textbf{x},$\beta$), where $\theta$ and $\phi$ encode the direction of the incoming particle, respectively the polar and azimuthal angles, and \textbf{x} is the vector representing the impact point on the mirror. These two dependencies are in very good approximation independent between each other and can be factored out. Thanks to this fact, by knowing the value that S$_{min}$ assumes at $\beta \approx$1 for every possible trajectory (S$_{min}|_{\beta=1}$), it is possible to calibrate the instrument for the general $\beta < 1$ case. 
This calibration was accomplished using a machine learning (ML) approach, based on a Boosted Decision Tree (BDT) technique \cite{bdt}.
\begin{figure}[t]
\centering
\includegraphics[width=0.49\textwidth]{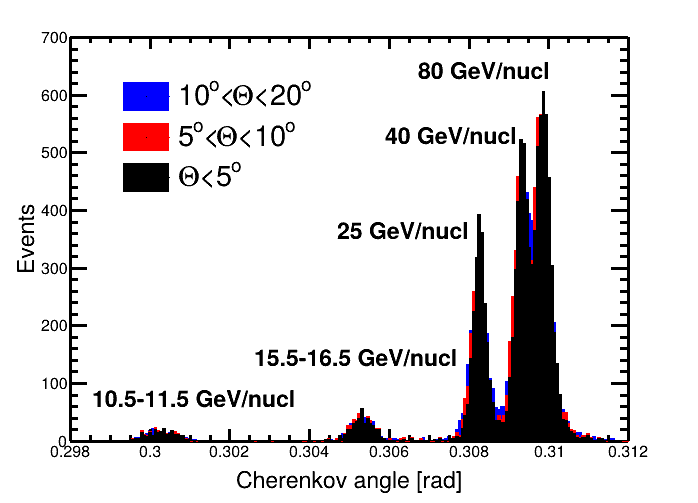}
\includegraphics[width=0.49\textwidth]{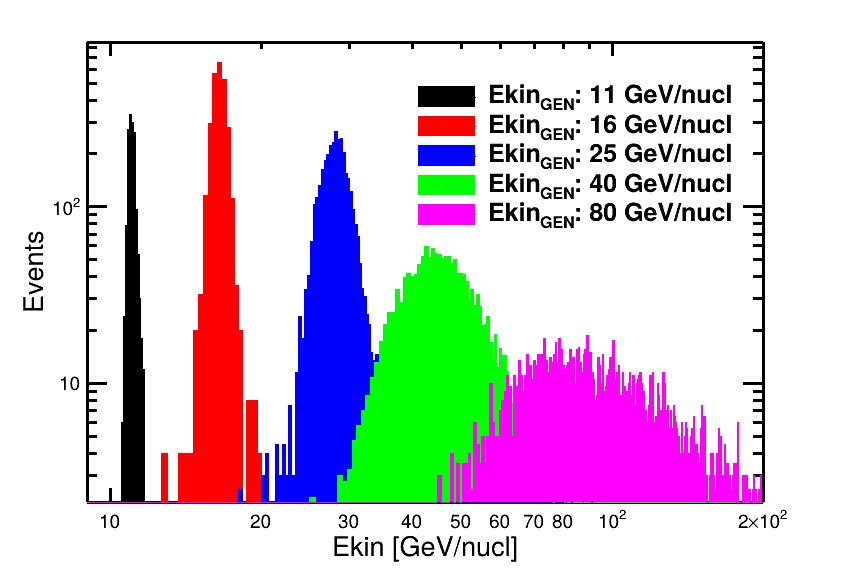}
\caption{Top: Distributions of reconstructed Cherenkov angles for samples of p and D generated at different energies with generic inclinations $\Theta$ [0$^\circ$-20$^\circ$] and impact points on the radiator. Different colors represent different intervals of inclination of the incoming particle.
Bottom: Distribution of measured $E_k$ for mono-energetic samples of p and D generated with generic inclinations $\Theta$ [0$^\circ$-20$^\circ$] and impact points on the radiator. Different colors represent different particle energies.  }
\label{fig:richperf}
\end{figure}

A BDT regression model was trained using the ROOT TMVA Toolkit \cite{tmva} on a mixture of $\beta \sim 1$ ($E_k = 200$ GeV/nucl.) $Z = 1$ particles, specifically protons and deuterons. The particles were uniformly generated from the top of the radiator plane with a continuous distribution of inclination angles relative to the detection plane.  

To construct the training set, the Cherenkov ellipses produced by these particles were fitted using the procedure described above, and for each event, the parameter S$_{min}|_{\beta=1}$ ($\theta$,$\phi$,\textbf{x}) was estimated. The BDT was trained to model the relationship between S$_{min}|_{\beta=1}$ and the particle's trajectory and to predict an estimated value S* that targets S$_{min}|_{\beta=1}$ ($\theta$,$\phi$,\textbf{x}) while using a set of input variables that are maximally independent of the particle's velocity.  

These input variables, derived from the particle trajectory reconstructed using the first two silicon layers of the MSIS, include the impact point on the radiator, the track inclination, and the incidence angle on the spherical mirror. The estimated \( S^* \) value is then used to calibrate the measurement at arbitrary velocity and inclination.  

In the approximation where the velocity dependence and the angular dependence G($\theta$,$\phi$,\textbf{x}) are disentangled, the measured minor axis can be expressed as  
 $$
 \mathrm{S_{min} (\theta,\phi,\textbf{x},\beta) = G(\theta,\phi,\textbf{x})\times R_{0}(\beta)}
 $$ 
 
 where \( R_{0}(\beta) \) is the radius of the Cherenkov ring for perpendicular tracks, unaffected by mirror-induced distortions. Since  

$$
\mathrm{S^* \approx S_{\min} |_{\beta=1} (\theta, \phi, \mathbf{x}) = G(\theta, \phi, \mathbf{x}) \times R_{0}(\beta=1)},
$$

the corresponding value of  R$_{0}$($\beta$) can be obtained as  

$$
\mathrm{R_{0}(\beta) = \left( \frac{S_{\min}(\theta, \phi, \mathbf{x}, \beta)}{S^*} \right) \times R_{0}(\beta=1)}.
$$

Given the general relation  R$_{0}$($\beta$) = d$_M \cdot \arctan(\Theta_c)$,

it is possible to obtain an estimate of the Cherenkov angle $\Theta_c$ that is independent of the particle trajectory.

Figure \ref{fig:bdtperf} illustrates the performance of this procedure on an independent test dataset simulated under conditions identical to those of the training set. The residuals between the value S* predicted by the BDT and the target variable S$_{min}|_{\beta=1}$ on the test dataset exhibit a Gaussian-like distribution. A Gaussian fit to the core of this distribution yields $\sigma_{BDT} = 3.44 \times 10^{-4}$, corresponding to an uncertainty on S* of 54.4 $\mu$m. This value is comparable to the resolution on the S$_{min}$ measurement. 

The final performance of the RICH system and its reconstruction algorithm is shown in Figure \ref{fig:richperf}, which depicts the results for protons and deuterons at various energies and generic impact point on the detector. No significant performance difference was found for the two particles. The total uncertainty observed in the Cherenkov angle measurement arises from multiple factors. These include the position resolution of the readout plane (see Appendix \ref{sec:sipm}), the imperfect focusing of the ellipse on the RICH plane (which dominates the $\sigma_{BDT}$ uncertainty on S*), and the contribution of the surface imperfections in the mirror. The resolution of the Cherenkov angle is nearly constant across the tested energy range, varying between 0.35 mrad and 0.45 mrad, and is compatible to the performance of other detectors with similar designs \cite{lhcbupgrade}. Thanks to the BDT-assisted calibration procedure, no significant dependence on the particle impact point and only minimal dependence on track inclination are shown in Figure \ref{fig:richperf} (top panel).
\begin{figure}[t]
\centering
\includegraphics[width=0.44\textwidth]{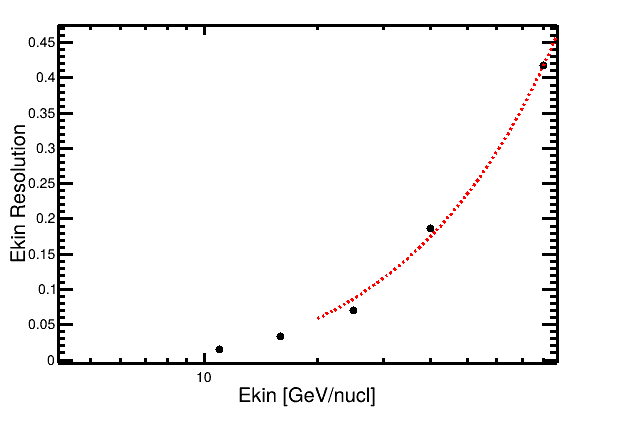}
\includegraphics[width=0.44\textwidth]{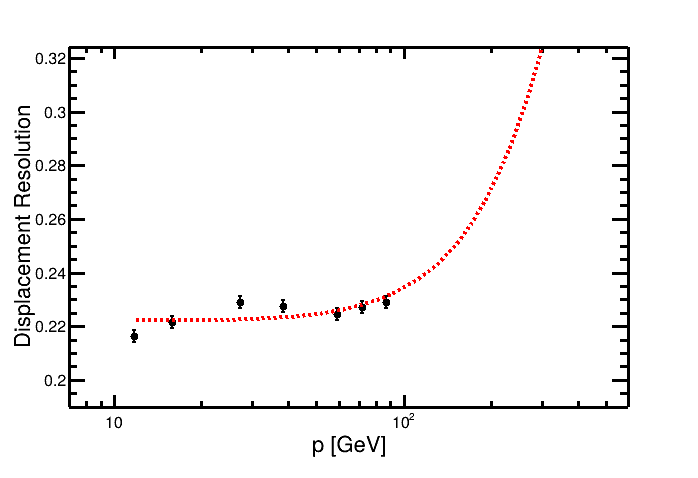}
\caption{Top: Resolution in the E$_k$ measurement by the DESMO RICH as a function of true E$_k$. 
A parabolic trend is superimposed to data to guide the eye (red dashed curve).
Bottom: Resolution of the measurement of the average displacement performed by DESMO for protons, as a function of particle momentum p, compared with the theoretical expectations discussed in the text for an optimized prototype with 9 stations (red dashed curve).}
\label{fig:resolutions}
\end{figure}

\begin{figure}[hb]
\centering
\includegraphics[width=0.44\textwidth]{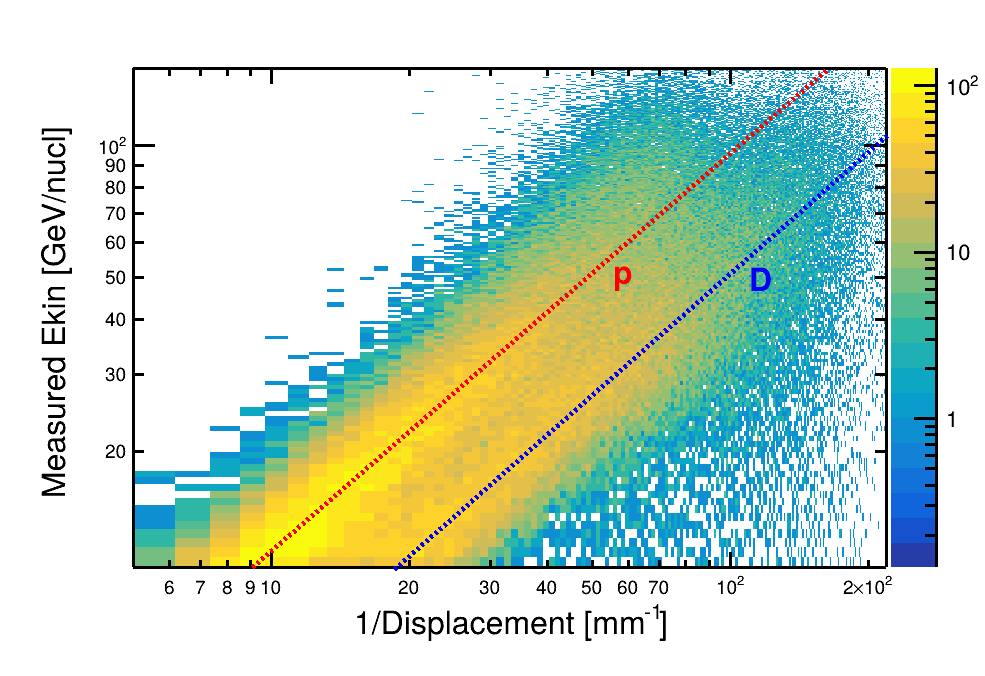}
\caption{The distribution of measured Ek versus measured inverse average displacement obtained by the DESMO pro- totype, for simulated events of D and p. The red and blue dotted lines represents respectively the theoretical expected behavior for protons and D nuclei.}
\label{fig:ekvsdisp}
\end{figure}
\section{DESMO expected performances}\label{perform}
The expected performance of DESMO were assessed with a simulation using the optimized design for cosmic D measurement. 
Monte Carlo simulations of p at different energies in the target E$_k$ energy range were used to estimate the resolutions that can be obtained in the measurements of average displacement and E$_k$. 
The results of this study are reported in Figure \ref{fig:resolutions}. The E$_k$ resolution below 20 GeV/nucl has values from 2 \% to 3 \%. At higher energies it shows a parabolic trend, up to 50 \% at 100 GeV/nucl. 
To minimize the bin-to-bin migration we divide the target energy range in 6 bins of increasing width, because of the degradation of the E$_k$ resolution. 
The resolution in the displacement measurement as a function of proton momentum is compared to the theoretical trend obtained combining the average displacement resolution calculated in Appendix \ref{sec:optim} (Equation \ref{eq:msformul} and Equation \ref{resoteo}).  The linearly rising trend at high energy, which falls outside the target energy range of DESMO, is where the overall displacement resolution starts to be importantly limited by the spatial resolution $\sigma_x$ of the detection plane. Consequently, the displacement resolution results in the range from 21\% to 23\%.

\begin{figure*}[ht]
\centering
\includegraphics[width=\textwidth]{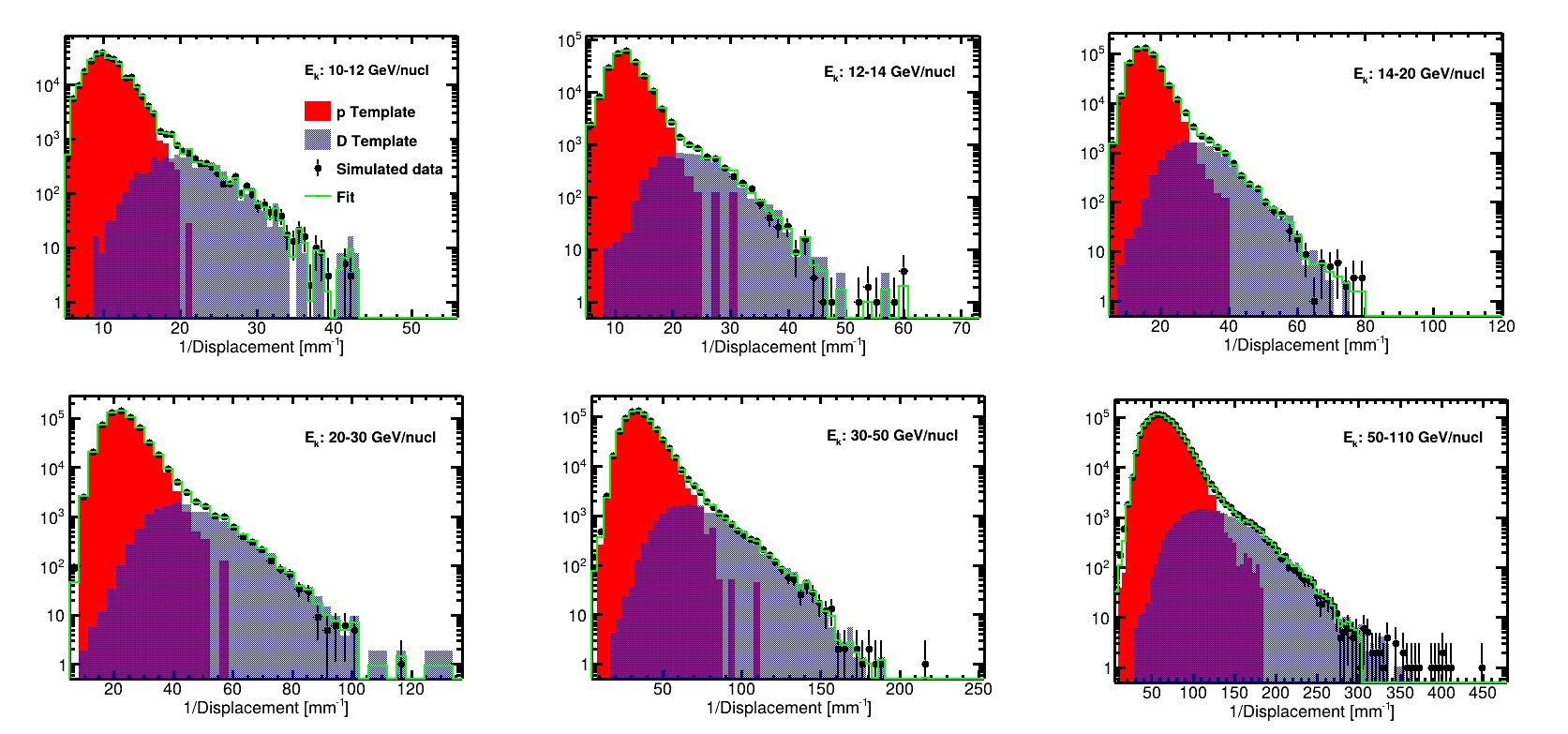}
\caption{Template fits of the inverse displacement distributions obtained from 60 days simulated cosmic Hydrogen flux, using the hypothesis of D abundance discussed in the text, in the six E$_k$ intervals measurable by DESMO. }
\label{fig:fit}
\end{figure*}
\begin{figure}[ht]
\centering
\includegraphics[width=0.5\textwidth]{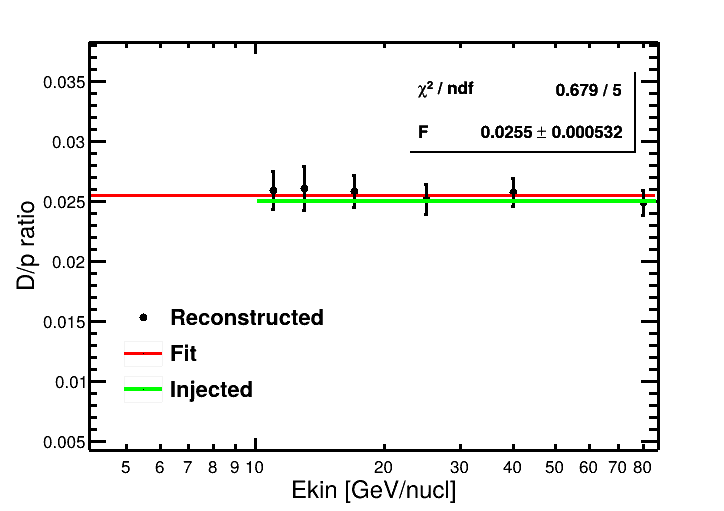}
\caption{Reconstructed D/p abundance detectable by a 60 days DESMO experiment on a simulated cosmic Hydrogen flux compared with the D fraction injected in the simulation (green solid line). Error bars represent statistical uncertainty from the template fit. A constant fit (red continuous line) on the reconstructed data yields a ratio F compatible with the injected one.}
\label{fig:fitresult}
\end{figure}
To evaluate the D/p separation performance, a GEANT simulation of the apparatus was performed with equal number D and p particles generated in an uniform E$_k$ distribution in the target energy range and different incidence angles in the range 0$^{\circ}$-20$^{\circ}$. The separation between the two populations can be observed in Figure \ref{fig:ekvsdisp}. 
The inverse displacement 1/d, measured in mm$^{-1}$ and proportional to particle momentum, was chosen as the separation variable for the analysis. This allows to compress the distribution of p and expand the distribution of the less abundant D, making the separation between the two when using realistic abundances more visible.
A similar simulation, with D and p events generated with an isotropic angular distribution from a square surface of 1$\times$1 m$^2$ placed above the detector, was used to estimate the detector acceptance. Only particles which create a Cherenkov reconstructed ring with the quality requirements discussed in Section \ref{RICHsection} and with a valid displacement measurement obtained by all of the  measuring stations are accepted. 

The angular view of the detector design is limited and it is possible to measure particles with $\sim$ 18$^\circ$ maximum inclination. This, combined with the efficiency of the selections described above, results in a almost constant acceptance of $\sim$ 40 cm$^2$ sr in the overall energy range.

Finally, the capabilities of the DESMO detector in D/p separation were evaluated simulating a realistic cosmic mixture of D and p, with a power-law spectrum in E$_k$ ($\propto$ E$_k^{-2.7}$) and constant D abundance of 2.5 \% in 10-100 GeV/nucl E$_k$ range. The total number of simulated events was chosen taking into account the total statistic attainable with the acceptance of the DESMO detector during a total exposure time of 60 days, achievable with a couple of balloon flights of 30 days each.
The simulated events passing the quality selection were collected in the 6 measured E$_k$ bins of the analysis and for each bin the 1/d distribution was evaluated. The distributions were fitted using p and D templates obtained from independent simulations using the ROOT MINUIT2 \cite{Hatlo:2005cj} minimization algorithm.
The results of the fit procedure for every E$_k$ bin are shown in Figure \ref{fig:fit}. As can be seen, with the simulated isotopic abundances, the p distribution alone is never sufficient to explain the observed distribution in simulated data. The D shoulder is always visible and it is possible to extract the D/p fraction from each of the bins.
The measured D/p fraction from this simulation is shown in Figure \ref{fig:fitresult}. As can be seen, although with a small systematic difference compatible with fit uncertainty, DESMO is able to measure the D relative abundance in good agreement with the injected ratio across the whole energy range. The results of these evaluation show a precision of the order of 5-8 \% on the D/p ratio, taking in account only the statistical uncertainty from the fit.
The precise evaluation of the systematic uncertainty given by the fragmentation of D and heavier nuclei such as $\mathrm{^4He}$ in (a) the residual atmosphere above the detector and (b) in the detector itself is beyond the purpose of this work. Uncertainty (a) is expected to be in the 2-5\% range \cite{Wang_2002}, and uncertainty (b)  to be below 5\% \cite{AMS2024}. These uncertainty estimations are a clear indication that a D abundance of the order of 1-2\% in the cosmic Hydrogen can be significantly detected by DESMO at energies up to 100 GeV/nucl. 
All the aforementioned studies demonstrate the effectiveness of DESMO on addressing his primary physics target. In addition, its measurement technique could be easily adapted to address other interesting cosmic-ray measurements. In Appendix \ref{scenarios}, we outline some of the most promising ones, accessible by DESMO directly or by upgraded versions of its design.

\section{Conclusion}

Studying the flux of cosmic-ray Deuterium (D) is a way to probe the mechanisms of cosmic-ray (CR) propagation and offers a unique insight into several topics: the investigation of the spallation processes, the composition of the interstellar medium (ISM), and the physical laws governing the CR diffusion. Deuterium, compared to heavier secondary species, with its light mass, enables to study the CR propagation over larger galactic distances. Additionally, its production from light primaries like $^4$He provides a complementary perspective to heavier secondary-to-primary ratios such as B/C.

The recent observations from the AMS02 experiment questioned our understating on these matters, challenging the traditional view of Deuterium as a purely secondary species. Its high-energy spectrum appears significantly harder than expected, resembling the one of protons, but deviating notably from the secondary $\mathrm{^3He}$. This suggests the existence of a primary-like Deuterium component, raising fundamental questions about CR propagation and about the universality postulate of CR diffusion. Understanding these deviations has implications not only with regards of cosmic-ray propagation, but also for indirect Dark Matter searches, since it helps to constrain the $\bar{D}$, a promising and nearly background-free channel.

In this work, we scrutinized the current knowledge about high-energy Deuterium, derived from the available experimental measurements and theoretical models. The emerging picture is complex, and its interpretation remains debated. Despite the precision measurements provided by AMS02 up to 21 GV, which hint toward a primary-like component, recent high-profile theoretical studies have shown that it is possible to explain this within a secondary framework. However, the large uncertainties in the nuclear cross sections and diffusion parameters still leave room for discussion on the theoretical side. Moreover, independent experimental results from SOKOL and CAPRICE98 show a significantly high Deuterium flux at high rigidities, which is difficult to contextualize within a secondary production scenario. 

To shade light on this topic an high-precision flux measurement, beyond the currently explored energy range, is required. This challenge could be addressed with DESMO (DEuterio con Scattering MultiplO), a balloon-borne experiment designed to measure the deuteron flux in the 20–200 GV rigidity range. DESMO leverages on a new experimental technique for the measurement of cosmic deuteron (D) flux at high energies. 
Matching stringent requirements on cost and complexity, it was designed to be operated as a particularly lightweight ($\sim$ 90 kg), relatively small (60$\times$60$\times$150 cm$^3$ ) and robust balloon-borne experiment, focused to explore the isotopic composition of the CR Hydrogen with a precision never reached before.
In this work we showed the performance of DESMO in distinguishing D, achieving a few percent level precision in the measurement of the D/p fraction, using a template-fit approach. Such performance, given the DESMO acceptance of 40 cm$^2$ sr, is attainable with a limited exposure time (60 days). The proposed DESMO design meets all feasibility requirements described in the paper and could enable a fundamental measurement capable of revealing the true nature of cosmic-ray Deuterium.

\section*{Acknowledgements}
We would like to express our gratitude to Dr. Francesco Nozzoli for the insightful discussions and for the bibliographic support. We also want to thank the referee for the useful comments and suggestions. 

\newpage
\appendix

\section{Different scientific scenarios}\label{scenarios}

Beyond delivering a deuteron flux measurement, DESMO also serves as a proof of concept for its novel measurement technique, which could be adopted for future applications.

In this section, following the same approach that led the design of DESMO, we outline various physics measurements achievable with minimal modifications to DESMO's design, or in the last most ambitious case, with the introduction of an additional module:

\begin{itemize}
\item \textbf{Gamma ray burst detection:}
High-energy gamma-rays can convert in the target material of DESMO, producing electron-positron pairs detectable by the following PPT modules. 
Electron-positron pair can also be easily measured by the DESMO BGO targets, which can effectively act as an electromagnetic calorimeter. 
The segmentation of the BGO targets and the signals of the silicon pixels would allow a 3D reconstruction of the electromagnetic shower and its distinction from hadronic ones. The absence of a signal in RICH will provide rejection against electron background and further anti-coincidence for hadrons.
Only the elaboration of advanced track analysis techniques are required to address this measurement.

\item \textbf{Low-Energy anti-protons flux measurement:} By precisely balancing the amount of target material of DESMO, low-energy anti-proton annihilation can be induced. The resulting pion star from $\bar{p}$ annihilation is easily detectable by tracking planes of DESMO, providing a signature for anti-protons below ~0.2 GeV/nucl.

\item \textbf{Cosmic $\bar{D}$ detection:} 
Detecting even a few confirmed anti-deuteron (\(\bar{D}\)) events would be a groundbreaking discovery in cosmic-ray physics, as it would provide strong evidence of annihilating DM in the galactic halo \cite{antid_1,antid_2}. The expected \(\bar{D}\) signal, concentrated at energies below 1 GeV/nucl, is extremely rare, with fluxes below \(10^{-5}\) m\(^2\) sec sr GeV/nucl \cite{Brauninger:2009pe}, and faces significant background interference from anti-protons (\(\bar{p}\)).
Current and future large-scale experiments like AMS02 \cite{amssens} and ALADINO \cite{ALADINO} have the potential to detect these signals, but specialized detectors such as the General Anti-Particle Spectrometer (GAPS) are also in an advanced developmental phase. 
As shown in Section \ref{sec:lowen} of the Appendix, the MSIS detector can be optimized for low energy studies, and for such can be shortened by a factor up to three, increasing its acceptance by a factor of 10 or more. 
The DESMO design described in this work cannot perform nuclei/anti-nuclei discrimination, since neither MSIS nor RICH can distinguish charge sign of the incoming particle. With the addition of a new module to the design, a 35\(\times\)35\(\times\)35 cm\(^3\) BGO calorimeter (CALO) matter/anti-matter discrimination and further distinction of \(\bar{D}\) from \(\bar{p}\) can be achievable. Moreover, replacing BGO as the target material in MSIS with faster high density scintillating crystals like Gadolinium Aluminium Gallium Garnet (GAGG) could also enable time-of-flight (TOF) measurements to determine particle velocity. The RICH would provide further anti-coincidence for high energy events. This modified design, with a total mass under 400 kg (dominated by CALO), remains well-suited for long-duration high-altitude polar balloon flights. The $\bar{D}$ experimental signature mirrors that of a similar-energy deuteron in MSIS and RICH but features a significantly higher energy deposit in CALO due to annihilation energy release. The mass difference relative to anti-protons $\bar{p}$ results in smaller displacements in MSIS and greater energy release in CALO, enabling $\bar{p}/\bar{D}$ distinction.
\end{itemize}

We acknowledge that the proposed improvements to the DESMO detector and its applicability to other experimental scenarios discussed in this section (expecially regarding the cosmic $\bar{D}$ detection) require further detailed studies.

\section{Optimization of MSIS design}\label{sec:optim}

The performances of the MSIS detector described in Section \ref{MSIS} in distinguishing isotopes of the same velocity fundamentally depend on 4 design parameters:

\begin{itemize}
\item target width x;
\item distance between subsequent PPT modules S;
\item spacing between the silicon planes of the PPT modules L;
\item number of measuring stations N.
\end{itemize}

It is possible to attempt an optimization of the MSIS design starting given the design limits specified in Section \ref{concept} and using:

\begin{equation}
    \Theta_{MS} = \frac{13.6 \, \mathrm{MeV}}{\beta p} z \sqrt{x/X_0} \left[ 1 + 0.038 \ln \left( \frac{x}{X_0} \right) \right]
    \label{eq:msformul}
\end{equation}

where the average scattering angle, $\Theta_{MS}$, depends on the particle's charge charge $z$ and momentum $p$ and on the thickness $x$ of the material, with radiation length $X_0$ \cite{pdg}.

The distribution of trajectory deviations $\theta$ given by a single PPT module can be approximated to a gaussian distribution  with $\sigma=\theta_{MS}$. Since only positive deviations from the trajectory measured by the precedent PPT module are measured, the average measured can be expressed as:
\begin{equation*}
\langle \theta \rangle = \int_{0}^{\infty} \theta \, \frac{1}{\theta_{\text{MS}} \sqrt{2\pi}} e^{-\frac{\theta^2}{2\theta_{\text{MS}}^2}} \, d\theta = \theta_{\text{MS}} \sqrt{\frac{2}{\pi}}
\end{equation*}

similarly, one can obtain the RMS value of the deviation from:

\begin{equation*}
\langle \theta^2 \rangle = \int_{0}^{\infty} \theta^2 \, \frac{1}{\theta_{\text{MS}} \sqrt{2\pi}} e^{-\frac{\theta^2}{2\theta_{\text{MS}}^2}} \, d\theta = \theta_{\text{MS}}^2
\end{equation*}

and finally obtain the expected deviation distribution width $\sigma_{\theta}$ from the well known equation
\begin{equation*}
\sigma^2_{\theta}=\langle \theta^2 \rangle - \langle \theta \rangle^2
\end{equation*}
which represents the intrinsic uncertainty of the MS induced deviation measurement.

Such uncertainty needs to be combined with the spatial uncertainty $\sigma_x$ of the hit position measurement and to the uncertainty on the position of the track interpolation $\sigma_I=\sqrt{2}(S/L)\sigma_x$ to obtain the final expression for the resolution on the average displacement resolution measured by N subsequent stations:

\begin{equation}
\frac{\sigma_d}{d}=\frac{1}{\theta_{MS}\sqrt{N}}\sqrt{\sigma^2_\theta + \sigma^2_x(\frac{1}{S^2}+\frac{2}{L^2})}\label{resoteo}
\end{equation}

Being S and L bound by S+L=L$_{TOT}$/N, where L$_{TOT}$ is the total length of the MSIS detector, the Expression \ref{resoteo} has a minimum for L=S$\approx$L$_{TOT}$/(2N).
Besides L$_{TOT}$, Expression \ref{resoteo} has also a dependence on the width of the target hidden in the $\theta_{MS}$ parameter, but the limits on the overall mass and length of the total detector discussed in Section \ref{concept} bound both of them, leaving only the number of stations N as a free parameter for the design optimization.

\begin{figure}[ht]
\centering
\includegraphics[width=0.45\textwidth]{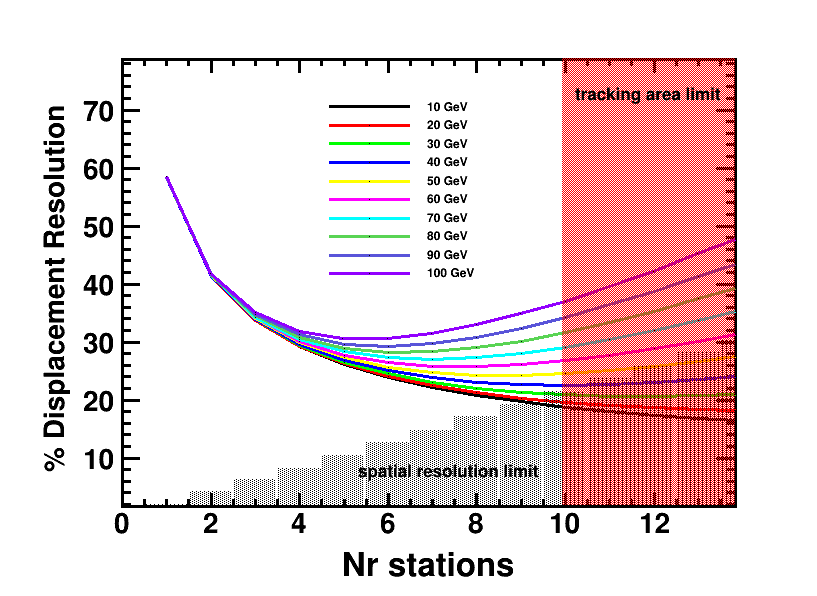}
\includegraphics[width=0.45\textwidth]{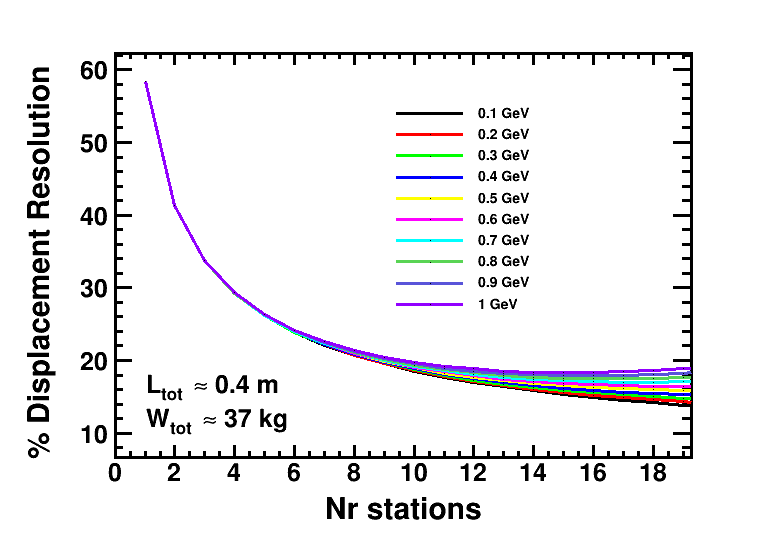}
\caption{Top: Resolution on average proton displacement measurement ($\sigma_{d}/d$)  expected (in \%) as a function of the number of PPT stations, considering an approximate total MSIS length of $\sim$ 1 m and a total weight of $\sim$ 90 kg. The red area corresponds to a total tracking area $\ge$ 2 m$^2$. The grey area corresponds to displacements at maximum E$_{k}$ below the spatial resolution of the silicon planes. 
Bottom: Resolution on average proton displacement measurement ($\sigma_{d}/d$)  expected (in \%)  as a function of the number of PPT stations, for different $E_k$ below 1 GeV/nucl, for a different MSIS prototype optimized for low energy studies. }
\label{fig:optim_highen}
\end{figure}

\subsection{Nominal use case (E $>$ 10 GeV/nucl)}
The predicted displacement resolution was calculated as a function of N and particle $E_k$, for ten different energies in the target energy range of the detector (10$\le E_k\le$100 GeV/nucl).
Figure \ref{fig:optim_highen} shows the results of this calculation. From the Figure it is visible that the curves at all the energies follow the general $\propto 1/\sqrt{N}$ trend. For higher energies and large values of N the impact of the spatial resolution becomes important, because since L + S is fixed, stations become shorter and shorter and a smaller displacement can be achieved (d $\propto$ S).
In particular, the same Figure shows also that designs with higher N approach the limit in wich the displacement at higher energy become comparable to the spatial resolution of the tracking planes.
A design with N = 8 was chosen as the best compromise that optimizes the displacement resolution across the whole energy range, still respecting the requirement on the total tracking area. 
This results in a final prototype with x=9 mm (included in S),  L=6 cm, S=6 cm and thus L$_{PPT}$=12 cm, for a total detector weight $\sim$ 90 kgm length $\sim$ 1 m, and tracking area below 1.5 m $^2$.

\subsection{Low energy case (E $\le$ 1 GeV/nucl)}\label{sec:lowen}

The same calculation can be used to devise a possible design optimized for the sub-GeV energies. In this case Equation \ref{eq:msformul} predicts much higher displacements, so the need for relatively high values of L$_{PPT}$ is less stringent. This fact can be exploited to obtain a shorter detector, effectively increasing the overall acceptance by a factor 10 or more at parity of all the other metrics. Likewise, also the thickness of the Pb targets can be reduced, obtaining a much lighter and compact design. 
As an example, Figure \ref{fig:optim_highen} shows the displacement resolution as a function of N for a prototype with total length and target thickness (thus total weight) scaled by a factor of 0.5 with respect to the configuration described for the nominal use case. As can be seen, despite the smaller lever arm and target thickness, the measured displacements remain relatively large, such as the contribution of $\sigma_{\theta}$ remains basically negligible up to very high N for all the energies below 1 GeV/nucl. 

As a result, for the low energy use case the most important limiting factor on the displacement resolution is the total tracking area.

\section{RICH readout plane}\label{sec:sipm}

Concerning the SiPM performance estimation discussed in this work, an averaged value of 50\% was chosen for Photon Detection Efficiency (PDE), accounting for sensor dead area, quantum efficiency in photon detection and probability to trigger an electronic signal \cite{lgsipm,AMBROSI2023168023}. We add a comparative study for the impact of different SiPM technologies (from well-established, to under development) on our estimation of the $E_{kin}$. The position resolution indeed is one of the dominant contribution to the uncertainty in the reconstruction of the ring properties in the RICH detector. For this study we identified three categories of position resolution $\sigma_x$, each of them achievable with a specific SiPM technology:
\begin{itemize}
    \item $\sigma_x \simeq 300$ $\mu$m, achievable with commercial SiPM produced by FBK, Hamamatsu, Broadcom. This SiPMs can achieve sub-millimeter resolution thanks to their pitch size, down to 1x1 mm$^2,$ and have integrated readout electronics already used in many space experiments \cite{Coutu_2024}.
    \item $\sigma_x \simeq 50-150$ $\mu$m, achievable with cutting-edge technologies, namely LG-SiPM \cite{lgsipm} and ultra-small SiPM \cite{YUE201538}. Being these technologies already tested for high-energy physics application, we took this case as a reference for the performance study shown in this work. 
    \item $\sigma_x \simeq 10-30$ $\mu$m, allowed by currently under development technologies based on digital single micro-cell readout as $d$SiPM \cite{dSiPM} or with Electron-Bombarded Active Pixel Sensors (EBAPS), which leverage on the MAPS position resolution \cite{EBAPS}.
    
\end{itemize}

\begin{figure}[ht]
\centering
\includegraphics[width=0.45\textwidth]{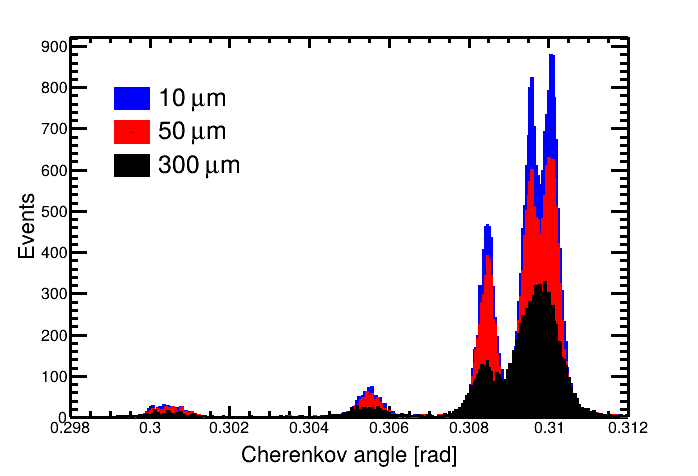}
\caption{Distributions of measured Cherenkov angles for samples of p and D selected as the ones shown in Figure \ref{fig:richperf} (top panel), reconstructed considering three different hypothesis of RICH readout plane resolution, as discussed in Appendix \ref{sec:sipm}.}
\label{fig:richnoise}
\end{figure}

Figure \ref{fig:richnoise} illustrates the impact of different sensor options on the RICH detector’s performance, showing the distribution of reconstructed Cherenkov angles for the D and p test sample discussed in Section \ref{RICHsection} (see also Figure \ref{fig:richperf}). In all cases except the worst-case scenario, the resolution is sufficient to distinguish the two highest energy populations (corresponding to true $E_{kin}$ of 40 GeV/nucl and 80 GeV/nucl). In this last case, the DESMO energy range is constrained to below 50 GeV/nucl. 

\begin{figure}[hb]
\centering
\includegraphics[width=0.49\textwidth]{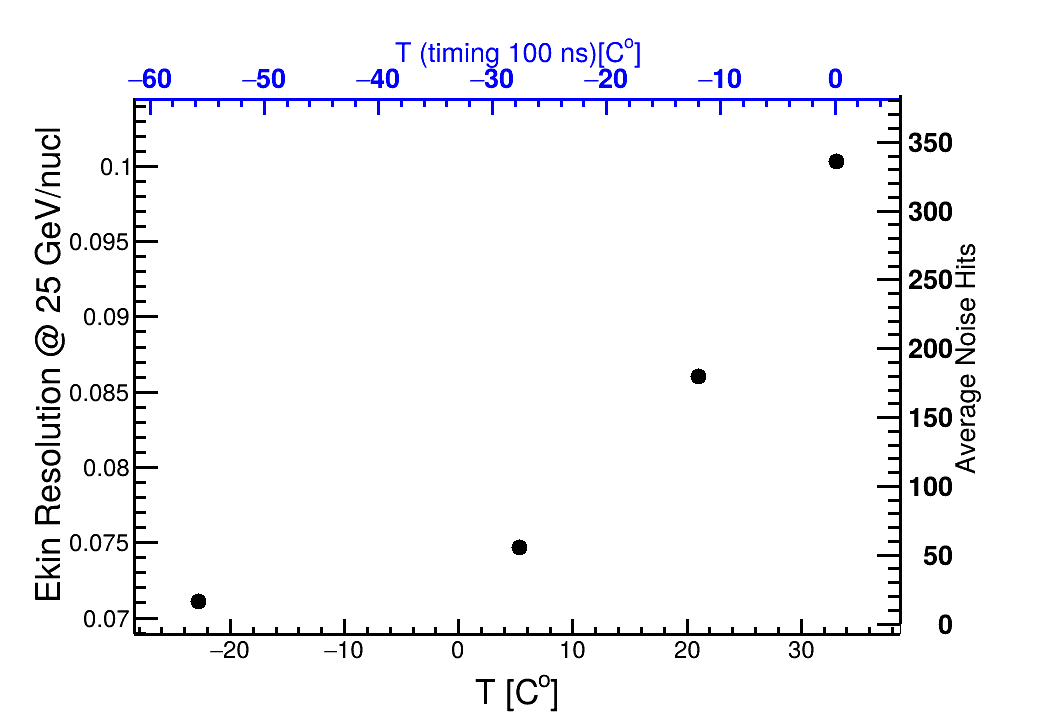}
\caption{$E_{kin}$ resolution of RICH as a function of operating temperature of DESMO, assuming a time window of 10 ns (100 ns - top blue axis) for the selection of Cherenkov hits. Right vertical axis represents the corresponding average number of DRC selected in the time window.}
\label{fig:figtemp}
\end{figure}

For all the aforementioned options (except the EBAPS) the issue of dark rate counts (DRC) must be taken into account. As mentioned in the paper, thermal control of the ballon and of the instrumentation on board (mirror and pixel detectors) will be in place, allowing for thermal stabilization at approximately 0$^\circ$. This feature, together with filtering techniques based on gating around the particle time of arrival \cite{Mazziotta_2025}, allows to mitigate the impact of DRC on the RICH measurement. A study of the mitigation of the impact of DRC is summarized in Figure \ref{fig:figtemp}, which shows the behaviour of measured resolution in $E_{kin}$ as a function of the operating temperature and therefore of the number of noisy hits (top x-axis). 

We assumed that commonly the DRC halves every 10$^\circ$ starting from a nominal value of 60 kcps/mm$^2$ (typical value at room temperature for LG-SiPMs) \cite{lgsipm}. With a typical time resolution for hit filtering of 10 ns \cite{helixprc,Mazziotta_2025} and an operating temperature of 0$^\circ$, the average contribution from fake hits is estimated to be approximately 50 hits. Nevertheless, Figure \ref{fig:figtemp} shows that these hits do not significantly impact the measurement, thanks to the fitting procedure, allowing DESMO to feature a $E_{kin}$ resolution at 25 GeV/nucl of 7\%. The Figure also shows, that a less precise time window (i.e 100 ns) would still result in a limited degradation of the performance provided an operating temperature of -15 C$^{\circ}$.

\bibliographystyle{unsrt}
\bibliography{aipsamp}

\begin{thebibliography}{10}

\bibitem{coste}
{Coste, B.}, {Derome, L.}, {Maurin, D.}, and {Putze, A.}
\newblock Constraining galactic cosmic-ray parameters with z $\le$ 2 nuclei.
\newblock {\em A\&A}, 539:A88, 2012.

\bibitem{PhysRevC.98.034611}
Yoann G\'enolini, David Maurin, Igor~V. Moskalenko, and Michael Unger.
\newblock Current status and desired precision of the isotopic production cross
  sections relevant to astrophysics of cosmic rays: Li, be, b, c, and n.
\newblock {\em Phys. Rev. C}, 98:034611, Sep 2018.

\bibitem{mosk}
Andrew~W. Strong and Igor~V. Moskalenko.
\newblock Propagation of cosmic-ray nucleons in the galaxy.
\newblock {\em The Astrophysical Journal}, 509(1):212, dec 1998.

\bibitem{Tomassetti_2017}
Nicola Tomassetti and Jie Feng.
\newblock The curious case of high-energy deuterons in galactic cosmic rays.
\newblock {\em The Astrophysical Journal Letters}, 835(2):L26, jan 2017.

\bibitem{PhysRevD.96.103005}
Nicola Tomassetti.
\newblock Solar and nuclear physics uncertainties in cosmic-ray propagation.
\newblock {\em Phys. Rev. D}, 96:103005, Nov 2017.

\bibitem{AMS2024}
M.~Aguilar, B.~Alpat, G.~Ambrosi, H.~Anderson, L.~Arruda, N.~Attig, C.~Bagwell,
  F.~Barao, M.~Barbanera, L.~Barrin, A.~Bartoloni, R.~Battiston, A.~Bayyari,
  N.~Belyaev, B.~Bertucci, V.~Bindi, K.~Bollweg, J.~Bolster, M.~Borchiellini,
  B.~Borgia, M.~J. Boschini, M.~Bourquin, C.~Brugnoni, J.~Burger, W.~J. Burger,
  X.~D. Cai, M.~Capell, J.~Casaus, G.~Castellini, F.~Cervelli, Y.~H. Chang,
  G.~M. Chen, G.~R. Chen, H.~Chen, H.~S. Chen, Y.~Chen, L.~Cheng, H.~Y. Chou,
  S.~Chouridou, V.~Choutko, C.~H. Chung, C.~Clark, G.~Coignet, C.~Consolandi,
  A.~Contin, C.~Corti, Z.~Cui, K.~Dadzie, F.~D'Angelo, A.~Dass, C.~Delgado,
  S.~Della~Torre, M.~B. Demirk\"oz, L.~Derome, S.~Di~Falco, V.~Di~Felice,
  C.~D\'{\i}az, F.~Dimiccoli, P.~von Doetinchem, F.~Dong, F.~Donnini,
  M.~Duranti, A.~Egorov, A.~Eline, F.~Faldi, D.~Fehr, J.~Feng, E.~Fiandrini,
  P.~Fisher, V.~Formato, C.~G\'amez, R.~J. Garc\'{\i}a-L\'opez, C.~Gargiulo,
  H.~Gast, M.~Gervasi, F.~Giovacchini, D.~M. G\'omez-Coral, J.~Gong, D.~Grandi,
  M.~Graziani, A.~N. Guracho, S.~Haino, K.~C. Han, R.~K. Hashmani, Z.~H. He,
  B.~Heber, T.~H. Hsieh, J.~Y. Hu, B.~W. Huang, M.~Ionica, M.~Incagli, Yi~Jia,
  H.~Jinchi, G.~Karag\"oz, S.~Khan, B.~Khiali, Th. Kirn, A.~P. Klipfel,
  O.~Kounina, A.~Kounine, V.~Koutsenko, D.~Krasnopevtsev, A.~Kuhlman,
  A.~Kulemzin, G.~La~Vacca, E.~Laudi, G.~Laurenti, G.~LaVecchia, I.~Lazzizzera,
  H.~T. Lee, S.~C. Lee, H.~L. Li, J.~Q. Li, M.~Li, M.~Li, Q.~Li, Q.~Li, Q.~Y.
  Li, S.~Li, S.~L. Li, J.~H. Li, Z.~H. Li, M.~J. Liang, P.~Liao, C.~H. Lin,
  T.~Lippert, J.~H. Liu, P.~C. Liu, S.~Q. Lu, Y.~S. Lu, K.~Luebelsmeyer, J.~Z.
  Luo, Q.~Luo, S.~D. Luo, Xi~Luo, C.~Ma\~n\'a, J.~Mar\'{\i}n, J.~Marquardt,
  G.~Mart\'{\i}nez, N.~Masi, D.~Maurin, T.~Medvedeva, A.~Menchaca-Rocha,
  Q.~Meng, V.~V. Mikhailov, M.~Molero, P.~Mott, L.~Mussolin, Y.~Najafi Jozani,
  R.~Nicolaidis, N.~Nikonov, F.~Nozzoli, J.~Ocampo-Peleteiro, A.~Oliva,
  M.~Orcinha, F.~Palmonari, M.~Paniccia, A.~Pashnin, M.~Pauluzzi, S.~Pensotti,
  P.~Pietzcker, V.~Plyaskin, S.~Poluianov, D.~Prid\"ohl, Z.~Y. Qu, L.~Quadrani,
  P.~G. Rancoita, D.~Rapin, A.~Reina Conde, E.~Robyn,
  I.~Rodr\'{\i}guez-Garc\'{\i}a, L.~Romaneehsen, F.~Rossi, A.~Rozhkov,
  D.~Rozza, R.~Sagdeev, E.~Savin, S.~Schael, A.~Schultz~von Dratzig,
  G.~Schwering, E.~S. Seo, B.~S. Shan, A.~Shukla, T.~Siedenburg, G.~Silvestre,
  J.~W. Song, X.~J. Song, R.~Sonnabend, L.~Strigari, T.~Su, Q.~Sun, Z.~T. Sun,
  L.~Tabarroni, M.~Tacconi, Z.~C. Tang, J.~Tian, Y.~Tian, Samuel C.~C. Ting,
  S.~M. Ting, N.~Tomassetti, J.~Torsti, T.~Urban, I.~Usoskin, V.~Vagelli,
  R.~Vainio, M.~Valencia-Otero, E.~Valente, E.~Valtonen, M.~V\'azquez~Acosta,
  M.~Vecchi, M.~Velasco, C.~X. Wang, L.~Wang, L.~Q. Wang, N.~H. Wang, Q.~L.
  Wang, S.~Wang, X.~Wang, Z.~M. Wang, J.~Wei, Z.~L. Weng, H.~Wu, Y.~Wu, Z.~B.
  Wu, J.~N. Xiao, R.~Q. Xiong, X.~Z. Xiong, W.~Xu, Q.~Yan, H.~T. Yang, Y.~Yang,
  A.~Yelland, H.~Yi, Y.~H. You, Y.~M. Yu, Z.~Q. Yu, C.~Zhang, F.~Z. Zhang,
  J.~Zhang, J.~H. Zhang, Z.~Zhang, P.~W. Zhao, C.~Zheng, Z.~M. Zheng, H.~L.
  Zhuang, V.~Zhukov, A.~Zichichi, and P.~Zuccon.
\newblock Properties of cosmic deuterons measured by the alpha magnetic
  spectrometer.
\newblock {\em Phys. Rev. Lett.}, 132:261001, Jun 2024.

\bibitem{Pamela}
O.~Adriani, G.~C. Barbarino, G.~A. Bazilevskaya, R.~Bellotti, M.~Boezio, E.~A.
  Bogomolov, M.~Bongi, V.~Bonvicini, S.~Bottai, A.~Bruno, F.~Cafagna,
  D.~Campana, P.~Carlson, M.~Casolino, G.~Castellini, C.~De Donato, C.~De
  Santis, N.~De Simone, V.~Di Felice, V.~Formato, A.~M. Galper, A.~V. Karelin,
  S.~V. Koldashov, S.~Koldobskiy, S.~Y. Krutkov, A.~N. Kvashnin, A.~Leonov,
  V.~Malakhov, L.~Marcelli, M.~Martucci, A.~G. Mayorov, W.~Menn, M~Merg{\`e},
  V.~V. Mikhailov, E.~Mocchiutti, A.~Monaco, N.~Mori, R.~Munini, G.~Osteria,
  F.~Palma, B.~Panico, P.~Papini, M.~Pearce, P.~Picozza, M.~Ricci, S.~B.
  Ricciarini, R.~Sarkar, V.~Scotti, M.~Simon, R.~Sparvoli, P.~Spillantini,
  Y.~I. Stozhkov, A.~Vacchi, E.~Vannuccini, G.~Vasilyev, S.~A. Voronov, Y.~T.
  Yurkin, G.~Zampa, and N.~Zampa.
\newblock Measurements of cosmic-ray hydrogen and helium isotopes with the
  pamela experiment.
\newblock {\em The Astrophysical Journal}, 818(1):68, feb 2016.

\bibitem{caprice}
P.~{Papini}, S.~{Piccardi}, P.~{Spillantini}, E.~{Vannuccini}, M.~{Ambriola},
  R.~{Bellotti}, F.~{Cafagna}, F.~{Ciacio}, M.~{Circella}, C.~N. {De Marzo},
  S.~{Bartalucci}, M.~{Ricci}, D.~{Bergstr{\"o}m}, P.~{Carlson}, T.~{Francke},
  P.~{Hansen}, E.~{Mocchiutti}, M.~{Boezio}, V.~{Bonvicini}, P.~{Schiavon},
  A.~{Vacchi}, N.~{Zampa}, U.~{Bravar}, S.~J. {Stochaj}, M.~{Casolino}, M.~P.
  {De Pascale}, A.~{Morselli}, P.~{Picozza}, R.~{Sparvoli}, M.~{Hof},
  J.~{Kremer}, W.~{Menn}, M.~{Simon}, J.~W. {Mitchell}, J.~F. {Ormes}, S.~A.
  {Stephens}, R.~E. {Streitmatter}, and M.~{Suffert}.
\newblock {High-Energy Deuteron Measurement with the CAPRICE98 Experiment}.
\newblock {\em \apj}, 615(1):259--274, November 2004.

\bibitem{sokol}
Andrey Turundaevskiy and Dmitry Podorozhnyi.
\newblock High energy deuterons in cosmic rays registered by the sokol
  satellite experiment.
\newblock {\em Advances in Space Research}, 59(1):496--501, 2017.

\bibitem{apprao1983}
K.~M.~V. Apparao.
\newblock {Flux of Cosmic Ray Deuterons with Rigidity Above 16.8 GV}.
\newblock In {\em Proceedings of the 13th International Conference on Cosmic
  Rays}, volume~1, page 126, Denver, Colorado, 1973.
\newblock {OG Sessions}.

\bibitem{Yuan_2024}
Qiang Yuan and Yi-Zhong Fan.
\newblock The ams-02 cosmic-ray deuteron flux is consistent with a secondary
  origin.
\newblock {\em The Astrophysical Journal Letters}, 974(1):L14, oct 2024.

\bibitem{antid_1}
F.~Donato, N.~Fornengo, and D.~Maurin.
\newblock Antideuteron fluxes from dark matter annihilation in diffusion
  models.
\newblock {\em Phys. Rev. D}, 78:043506, Aug 2008.

\bibitem{bbnuc}
Ryan~J. Cooke, Max Pettini, and Charles~C. Steidel.
\newblock One percent determination of the primordial deuterium abundance*.
\newblock {\em The Astrophysical Journal}, 855(2):102, mar 2018.

\bibitem{galprop}
A.E. Vladimirov, S.W. Digel, G.~Jóhannesson, P.F. Michelson, I.V. Moskalenko,
  P.L. Nolan, E.~Orlando, T.A. Porter, and A.W. Strong.
\newblock Galprop webrun: An internet-based service for calculating galactic
  cosmic ray propagation and associated photon emissions.
\newblock {\em Computer Physics Communications}, 182(5):1156--1161, 2011.

\bibitem{arx}
Xing-Jian Lv, Xiao-Jun Bi, Kun Fang, Peng-Fei Yin, and Meng-Jie Zhao.
\newblock Cosmic-ray deuteron excess from a primary component, 2024.

\bibitem{DampeP}
DAMPE Collaboration, Q.~An, R.~Asfandiyarov, P.~Azzarello, P.~Bernardini, X.~J.
  Bi, M.~S. Cai, J.~Chang, D.~Y. Chen, H.~F. Chen, J.~L. Chen, W.~Chen, M.~Y.
  Cui, T.~S. Cui, H.~T. Dai, A.~D’Amone, A.~De Benedittis, I.~De Mitri, M.~Di
  Santo, M.~Ding, T.~K. Dong, Y.~F. Dong, Z.~X. Dong, G.~Donvito, D.~Droz,
  J.~L. Duan, K.~K. Duan, D.~D’Urso, R.~R. Fan, Y.~Z. Fan, F.~Fang, C.~Q.
  Feng, L.~Feng, P.~Fusco, V.~Gallo, F.~J. Gan, M.~Gao, F.~Gargano, K.~Gong,
  Y.~Z. Gong, D.~Y. Guo, J.~H. Guo, X.~L. Guo, S.~X. Han, Y.~M. Hu, G.~S.
  Huang, X.~Y. Huang, Y.~Y. Huang, M.~Ionica, W.~Jiang, X.~Jin, J.~Kong, S.~J.
  Lei, S.~Li, W.~L. Li, X.~Li, X.~Q. Li, Y.~Li, Y.~F. Liang, Y.~M. Liang, N.~H.
  Liao, C.~M. Liu, H.~Liu, J.~Liu, S.~B. Liu, W.~Q. Liu, Y.~Liu, F.~Loparco,
  C.~N. Luo, M.~Ma, P.~X. Ma, S.~Y. Ma, T.~Ma, X.~Y. Ma, G.~Marsella, M.~N.
  Mazziotta, D.~Mo, X.~Y. Niu, X.~Pan, W.~X. Peng, X.~Y. Peng, R.~Qiao, J.~N.
  Rao, M.~M. Salinas, G.~Z. Shang, W.~H. Shen, Z.~Q. Shen, Z.~T. Shen, J.~X.
  Song, H.~Su, M.~Su, Z.~Y. Sun, A.~Surdo, X.~J. Teng, A.~Tykhonov, S.~Vitillo,
  C.~Wang, H.~Wang, H.~Y. Wang, J.~Z. Wang, L.~G. Wang, Q.~Wang, S.~Wang, X.~H.
  Wang, X.~L. Wang, Y.~F. Wang, Y.~P. Wang, Y.~Z. Wang, Z.~M. Wang, D.~M. Wei,
  J.~J. Wei, Y.~F. Wei, S.~C. Wen, D.~Wu, J.~Wu, L.~B. Wu, S.~S. Wu, X.~Wu,
  K.~Xi, Z.~Q. Xia, H.~T. Xu, Z.~H. Xu, Z.~L. Xu, Z.~Z. Xu, G.~F. Xue, H.~B.
  Yang, P.~Yang, Y.~Q. Yang, Z.~L. Yang, H.~J. Yao, Y.~H. Yu, Q.~Yuan, C.~Yue,
  J.~J. Zang, F.~Zhang, J.~Y. Zhang, J.~Z. Zhang, P.~F. Zhang, S.~X. Zhang,
  W.~Z. Zhang, Y.~Zhang, Y.~J. Zhang, Y.~L. Zhang, Y.~P. Zhang, Y.~Q. Zhang,
  Z.~Zhang, Z.~Y. Zhang, H.~Zhao, H.~Y. Zhao, X.~F. Zhao, C.~Y. Zhou, Y.~Zhou,
  X.~Zhu, Y.~Zhu, and S.~Zimmer.
\newblock Measurement of the cosmic ray proton spectrum from 40 gev to 100 tev
  with the dampe satellite.
\newblock {\em Science Advances}, 5(9):eaax3793, 2019.

\bibitem{PhysRevLett.119.251101}
M.~Aguilar, L.~Ali~Cavasonza, B.~Alpat, G.~Ambrosi, L.~Arruda, N.~Attig,
  S.~Aupetit, P.~Azzarello, A.~Bachlechner, F.~Barao, A.~Barrau, L.~Barrin,
  A.~Bartoloni, L.~Basara, S.~Ba\ifmmode \mbox{\c{s}}\else \c{s}\fi{}e\ifmmode
  \breve{g}\else \u{g}\fi{}mez-du Pree, M.~Battarbee, R.~Battiston, U.~Becker,
  M.~Behlmann, B.~Beischer, J.~Berdugo, B.~Bertucci, K.~F. Bindel, V.~Bindi,
  W.~de~Boer, K.~Bollweg, V.~Bonnivard, B.~Borgia, M.~J. Boschini, M.~Bourquin,
  E.~F. Bueno, J.~Burger, W.~J. Burger, F.~Cadoux, X.~D. Cai, M.~Capell,
  S.~Caroff, J.~Casaus, G.~Castellini, F.~Cervelli, M.~J. Chae, Y.~H. Chang,
  A.~I. Chen, G.~M. Chen, H.~S. Chen, L.~Cheng, H.~Y. Chou, E.~Choumilov,
  V.~Choutko, C.~H. Chung, C.~Clark, R.~Clavero, G.~Coignet, C.~Consolandi,
  A.~Contin, C.~Corti, W.~Creus, M.~Crispoltoni, Z.~Cui, K.~Dadzie, Y.~M. Dai,
  A.~Datta, C.~Delgado, S.~Della~Torre, O.~Demakov, M.~B. Demirk\"oz,
  L.~Derome, S.~Di~Falco, F.~Dimiccoli, C.~D\'{\i}az, P.~von Doetinchem,
  F.~Dong, F.~Donnini, M.~Duranti, D.~D'Urso, A.~Egorov, A.~Eline, T.~Eronen,
  J.~Feng, E.~Fiandrini, P.~Fisher, V.~Formato, Y.~Galaktionov, G.~Gallucci,
  R.~J. Garc\'{\i}a-L\'opez, C.~Gargiulo, H.~Gast, I.~Gebauer, M.~Gervasi,
  A.~Ghelfi, F.~Giovacchini, D.~M. G\'omez-Coral, J.~Gong, C.~Goy, V.~Grabski,
  D.~Grandi, M.~Graziani, K.~H. Guo, S.~Haino, K.~C. Han, Z.~H. He, M.~Heil,
  J.~Hoffman, T.~H. Hsieh, H.~Huang, Z.~C. Huang, C.~Huh, M.~Incagli,
  M.~Ionica, W.~Y. Jang, Yi~Jia, H.~Jinchi, S.~C. Kang, K.~Kanishev, B.~Khiali,
  G.~N. Kim, K.~S. Kim, Th. Kirn, C.~Konak, O.~Kounina, A.~Kounine,
  V.~Koutsenko, A.~Kulemzin, G.~La~Vacca, E.~Laudi, G.~Laurenti, I.~Lazzizzera,
  A.~Lebedev, H.~T. Lee, S.~C. Lee, C.~Leluc, H.~S. Li, J.~Q. Li, Q.~Li, T.~X.
  Li, Y.~Li, Z.~H. Li, Z.~Y. Li, S.~Lim, C.~H. Lin, P.~Lipari, T.~Lippert,
  D.~Liu, Hu~Liu, V.~D. Lordello, S.~Q. Lu, Y.~S. Lu, K.~Luebelsmeyer, F.~Luo,
  J.~Z. Luo, S.~S. Lyu, F.~Machate, C.~Ma\~n\'a, J.~Mar\'{\i}n, T.~Martin,
  G.~Mart\'{\i}nez, N.~Masi, D.~Maurin, A.~Menchaca-Rocha, Q.~Meng, V.~M.
  Mikuni, D.~C. Mo, P.~Mott, T.~Nelson, J.~Q. Ni, N.~Nikonov, F.~Nozzoli,
  A.~Oliva, M.~Orcinha, F.~Palmonari, C.~Palomares, M.~Paniccia, M.~Pauluzzi,
  S.~Pensotti, C.~Perrina, H.~D. Phan, N.~Picot-Clemente, F.~Pilo,
  C.~Pizzolotto, V.~Plyaskin, M.~Pohl, V.~Poireau, L.~Quadrani, X.~M. Qi,
  X.~Qin, Z.~Y. Qu, T.~R\"aih\"a, P.~G. Rancoita, D.~Rapin, J.~S. Ricol,
  S.~Rosier-Lees, A.~Rozhkov, D.~Rozza, R.~Sagdeev, S.~Schael, S.~M. Schmidt,
  A.~Schulz~von Dratzig, G.~Schwering, E.~S. Seo, B.~S. Shan, J.~Y. Shi,
  T.~Siedenburg, D.~Son, J.~W. Song, M.~Tacconi, X.~W. Tang, Z.~C. Tang,
  D.~Tescaro, Samuel C.~C. Ting, S.~M. Ting, N.~Tomassetti, J.~Torsti,
  C.~T\"urko\ifmmode~\breve{g}\else \u{g}\fi{}lu, T.~Urban, V.~Vagelli,
  E.~Valente, E.~Valtonen, M.~V\'azquez~Acosta, M.~Vecchi, M.~Velasco, J.~P.
  Vialle, V.~Vitale, S.~Vitillo, L.~Q. Wang, N.~H. Wang, Q.~L. Wang, X.~Wang,
  X.~Q. Wang, Z.~X. Wang, C.~C. Wei, Z.~L. Weng, K.~Whitman, H.~Wu, X.~Wu,
  R.~Q. Xiong, W.~Xu, Q.~Yan, J.~Yang, M.~Yang, Y.~Yang, H.~Yi, Y.~J. Yu, Z.~Q.
  Yu, M.~Zannoni, S.~Zeissler, C.~Zhang, F.~Zhang, J.~Zhang, J.~H. Zhang, S.~W.
  Zhang, Z.~Zhang, Z.~M. Zheng, H.~L. Zhuang, V.~Zhukov, A.~Zichichi,
  N.~Zimmermann, and P.~Zuccon.
\newblock Observation of the identical rigidity dependence of he, c, and o
  cosmic rays at high rigidities by the alpha magnetic spectrometer on the
  international space station.
\newblock {\em Phys. Rev. Lett.}, 119:251101, Dec 2017.

\bibitem{DimiccoliFollega2024}
Francesco Dimiccoli and Francesco~Maria Follega.
\newblock A multiple scattering-based technique for isotopic identification in
  cosmic rays.
\newblock {\em Particles}, 7(2):477--488, 2024.

\bibitem{Agafonova_2012}
N~Agafonova, A~Aleksandrov, O~Altinok, A~Anokhina, S~Aoki, A~Ariga, T~Ariga,
  D~Autiero, A~Badertscher, A~Bagulya, A~Ben Dhahbi, A~Bertolin, M~Besnier,
  C~Bozza, T~Brugière, R~Brugnera, F~Brunet, G~Brunetti, S~Buontempo, A~Cazes,
  L~Chaussard, M~Chernyavskiy, V~Chiarella, A~Chukanov, N~D'Ambrosio, F~Dal
  Corso, G~De Lellis, P~del Amo~Sanchez, Y~Déclais, M~De Serio, F~Di Capua,
  A~Di Crescenzo, D~Di Ferdinando, N~Di Marco, S~Dmitrievski, M~Dracos,
  D~Duchesneau, S~Dusini, T~Dzhatdoev, J~Ebert, O~Egorov, R~Enikeev,
  A~Ereditato, L~S Esposito, J~Favier, T~Ferber, R~A Fini, D~Frekers, T~Fukuda,
  A~Garfagnini, G~Giacomelli, M~Giorgini, C~Göllnitz, J~Goldberg, D~Golubkov,
  L~Goncharova, Y~Gornushkin, G~Grella, F~Grianti, A~M Guler, C~Gustavino,
  C~Hagner, K~Hamada, T~Hara, M~Hierholzer, A~Hollnagel, K~Hoshino, M~Ieva,
  H~Ishida, K~Jakovcic, C~Jollet, F~Juget, M~Kamiscioglu, K~Kazuyama, S~H Kim,
  M~Kimura, N~Kitagawa, B~Klicek, J~Knuesel, K~Kodama, M~Komatsu, U~Kose,
  I~Kreslo, H~Kubota, C~Lazzaro, J~Lenkeit, I~Lippi, A~Ljubicic, A~Longhin,
  P~Loverre, G~Lutter, A~Malgin, G~Mandrioli, K~Manai, J~Marteau, T~Matsuo,
  V~Matveev, N~Mauri, E~Medinaceli, F~Meisel, A~Meregaglia, P~Migliozzi,
  S~Mikado, S~Miyamoto, P~Monacelli, K~Morishima, U~Moser, M~T Muciaccia,
  N~Naganawa, T~Naka, M~Nakamura, T~Nakano, D~Naumov, V~Nikitina, K~Niwa,
  Y~Nonoyama, S~Ogawa, N~Okateva, A~Olshevskiy, M~Paniccia, A~Paoloni, B~D
  Park, I~G Park, A~Pastore, L~Patrizii, E~Pennacchio, H~Pessard, K~Pretzl,
  V~Pilipenko, C~Pistillo, N~Polukhina, M~Pozzato, F~Pupilli, R~Rescigno,
  T~Roganova, H~Rokujo, G~Romano, G~Rosa, I~Rostovtseva, A~Rubbia, A~Russo,
  V~Ryasny, O~Ryazhskaya, O~Sato, Y~Sato, A~Schembri, W~Schmidt-Parzefall,
  H~Schroeder, L~Scotto Lavina, A~Sheshukov, H~Shibuya, G~Shoziyoev, S~Simone,
  M~Sioli, C~Sirignano, G~Sirri, J~S Song, M~Spinetti, L~Stanco, N~Starkov,
  M~Stipcevic, T~Strauss, P~Strolin, S~Takahashi, M~Tenti, F~Terranova,
  I~Tezuka, V~Tioukov, P~Tolun, A~Trabelsi, T~Tran, S~Tufanli, P~Vilain,
  M~Vladimirov, L~Votano, J~L Vuilleumier, G~Wilquet, B~Wonsak, V~Yakushev, C~S
  Yoon, T~Yoshioka, J~Yoshida, Y~Zaitsev, S~Zemskova, A~Zghiche, and
  R~Zimmermann.
\newblock Momentum measurement by the multiple coulomb scattering method in the
  opera lead-emulsion target.
\newblock {\em New Journal of Physics}, 14(1):013026, jan 2012.

\bibitem{geant4}
S.~Agostinelli et~al.
\newblock {GEANT4--a simulation toolkit}.
\newblock {\em Nucl. Instrum. Meth. A}, 506:250--303, 2003.

\bibitem{AGLIERIRINELLA2017583}
Gianluca {Aglieri Rinella}.
\newblock The alpide pixel sensor chip for the upgrade of the alice inner
  tracking system.
\newblock {\em Nuclear Instruments and Methods in Physics Research Section A:
  Accelerators, Spectrometers, Detectors and Associated Equipment},
  845:583--587, 2017.
\newblock Proceedings of the Vienna Conference on Instrumentation 2016.

\bibitem{ester}
Ester Ricci.
\newblock {A pixel based tracker for the HEPD-02 detector}.
\newblock {\em PoS}, ICRC2023:164, 2024.

\bibitem{savino}
Umberto Savino.
\newblock Expected performance of the high energy particle detector (hepd-02)
  tracking system on board of the second china seismo-electromagnetic
  satellite.
\newblock {\em Nuclear Instruments and Methods in Physics Research Section A:
  Accelerators, Spectrometers, Detectors and Associated Equipment},
  1063:169281, 03 2024.

\bibitem{bgo}
Mohammad Farukhi.
\newblock Bi4ge3o12 (bgo) - a scintillator replacement for nai(tl).
\newblock {\em MRS Proceedings}, 16, 01 2011.

\bibitem{MAGER2016434}
M.~Mager.
\newblock Alpide, the monolithic active pixel sensor for the alice its upgrade.
\newblock {\em Nuclear Instruments and Methods in Physics Research Section A:
  Accelerators, Spectrometers, Detectors and Associated Equipment},
  824:434--438, 2016.
\newblock Frontier Detectors for Frontier Physics: Proceedings of the 13th Pisa
  Meeting on Advanced Detectors.

\bibitem{Vuchkov2001}
Ivan~N. Vuchkov and Lidia~N. Boyadjieva.
\newblock {\em Statistical Methods for Data Analysis}, pages 14--95.
\newblock Springer Netherlands, Dordrecht, 2001.

\bibitem{viehhauser_weidberg_2024}
Georg Viehhauser and Tony Weidberg.
\newblock {\em Detectors in Particle Physics: A Modern Introduction}.
\newblock CRC Press, 1st edition, 2024.

\bibitem{PWang_1994}
P~Wang, A~Beck, W~Korner, H~Scheller, and J~Fricke.
\newblock Density and refractive index of silica aerogels after low- and
  high-temperature supercritical drying and thermal treatment.
\newblock {\em Journal of Physics D: Applied Physics}, 27(2):414, feb 1994.

\bibitem{Giovacchini:2023ixx}
F.~Giovacchini.
\newblock {The RICH detector of the AMS-02 experiment aboard the International
  Space Station}.
\newblock {\em Nucl. Instrum. Meth. A}, 1055:168434, 2023.

\bibitem{PEREIRA}
RUI PEREIRA and on~behalf of~the AMS RICH~collaboration.
\newblock {\em The RICH detector of the AMS-02 experiment: status and physics
  prospects}, pages 901--905.
\newblock World Scientific, 2008.

\bibitem{lgsipm}
Fabio Acerbi, Stefano Merzi, and Alberto Gola.
\newblock Position-sensitive silicon photomultiplier arrays with large-area and
  sub-millimeter resolution.
\newblock {\em Sensors}, 24(14), 2024.

\bibitem{helixprc}
Naram Park, P.~Allison, L.~Beaufore, Y.~Chen, Stephane Coutu, M.~Gebhard, Noah
  Green, Diego Hanna, H.B. Jeon, B.~Kunkler, M.~Lang, R.~Mbarek, K.~McBride,
  I.~Mognet, D.~Muller, S.~Nutter, Stephan O'Brien, and M.~Yu.
\newblock Cosmic-ray isotope measurements with helix.
\newblock page 091, 07 2021.

\bibitem{gaps}
C~J Hailey.
\newblock An indirect search for dark matter using antideuterons: the gaps
  experiment.
\newblock {\em New Journal of Physics}, 11(10):105022, oct 2009.

\bibitem{AMBROSI}
G.~Ambrosi, M.~Ambrosio, C.~Aramo, B.~Bertucci, E.~Bissaldi, M.~Bitossi,
  A.~Boiano, C.~Bonavolontà, M.~Capasso, A.~Circiello, L.~Consiglio,
  D.~Depaoli, F.~{Di Pierro}, L.~{Di Venere}, E.~Fiandrini, N.~Giglietto,
  F.~Giordano, S.~Incardona, M.~Ionica, F.~Licciulli, S.~Loporchio,
  G.~Marsella, V.~Masone, F.R. Pantaleo, R.~Paoletti, B.~Ruggiero,
  A.~Rugliancich, P.~Silvestrini, L.~Stiaccini, J.~Tasseva, L.~Tosti,
  G.~Tripodo, V.~Vagelli, and M.~Valentino.
\newblock High-density near-ultraviolet silicon photomultipliers:
  Characterization of photosensors for cherenkov light detection.
\newblock {\em Nuclear Instruments and Methods in Physics Research Section A:
  Accelerators, Spectrometers, Detectors and Associated Equipment},
  1049:168023, 2023.

\bibitem{iwasi}
A.~R. Altamura, A.~Di Mauro, E.~Nappi, N.~Nicassio, M.~van Emmerik, and
  G.~Volpe.
\newblock {Aerogel characterization for RICH applications}.
\newblock In {\em {9th International Workshop on Advances in Sensors and
  Interfaces}}, 6 2023.

\bibitem{CISBANI2003305}
E.~Cisbani, S.~Colilli, R.~Crateri, F.~Cusanno, R.~Fratoni, S.~Frullani,
  F.~Garibaldi, F.~Giuliani, M.~Gricia, M.~Iodice, R.~Iommi, M.~Lucentini,
  A.~Mostarda, L.~Pierangeli, F.~Santavenere, G.M. Urciuoli, R.~{De Leo},
  L.~Lagamba, E.~Nappi, A.~Braem, and P.~Vernin.
\newblock Light-weight spherical mirrors for cherenkov detectors.
\newblock {\em Nuclear Instruments and Methods in Physics Research Section A:
  Accelerators, Spectrometers, Detectors and Associated Equipment},
  496(2):305--314, 2003.

\bibitem{bdt}
Yann Coadou.
\newblock Boosted decision trees and applications.
\newblock {\em EPJ Web of Conferences}, 55:02004--, 07 2013.

\bibitem{tmva}
Jan Therhaag.
\newblock {TMVA Toolkit for multivariate data analysis in ROOT}.
\newblock {\em PoS}, ICHEP2010:510, 2010.

\bibitem{lhcbupgrade}
S.A. Wotton.
\newblock The lhcb rich upgrade for the high luminosity lhc era.
\newblock {\em Nuclear Instruments and Methods in Physics Research Section A:
  Accelerators, Spectrometers, Detectors and Associated Equipment},
  1058:168824, 2024.

\bibitem{Hatlo:2005cj}
M.~Hatlo, F.~James, P.~Mato, L.~Moneta, M.~Winkler, and A.~Zsenei.
\newblock {Developments of mathematical software libraries for the LHC
  experiments}.
\newblock {\em IEEE Trans. Nucl. Sci.}, 52:2818--2822, 2005.

\bibitem{Wang_2002}
J.~Z. Wang, E.~S. Seo, K.~Anraku, M.~Fujikawa, M.~Imori, T.~Maeno, N.~Matsui,
  H.~Matsunaga, M.~Motoki, S.~Orito, T.~Saeki, T.~Sanuki, I.~Ueda,
  K.~Yoshimura, Y.~Makida, J.~Suzuki, K.~Tanaka, A.~Yamamoto, T.~Yoshida,
  T.~Mitsui, H.~Matsumoto, M.~Nozaki, M.~Sasaki, J.~Mitchell, A.~Moiseev,
  J.~Ormes, R.~Streitmatter, J.~Nishimura, Y.~Yajima, and T.~Yamagami.
\newblock Measurement of cosmic-ray hydrogen and helium and their isotopic
  composition with the bess experiment.
\newblock {\em The Astrophysical Journal}, 564(1):244, jan 2002.

\bibitem{antid_2}
T.~Aramaki, S.~Boggs, S.~Bufalino, L.~Dal, P.~{von Doetinchem}, F.~Donato,
  N.~Fornengo, H.~Fuke, M.~Grefe, C.~Hailey, B.~Hamilton, A.~Ibarra,
  J.~Mitchell, I.~Mognet, R.A. Ong, R.~Pereira, K.~Perez, A.~Putze, A.~Raklev,
  P.~Salati, M.~Sasaki, G.~Tarle, A.~Urbano, A.~Vittino, S.~Wild, W.~Xue, and
  K.~Yoshimura.
\newblock Review of the theoretical and experimental status of dark matter
  identification with cosmic-ray antideuterons.
\newblock {\em Physics Reports}, 618:1--37, 2016.
\newblock Review of the theoretical and experimental status of dark matter
  identification with cosmic-ray antideuterons.

\bibitem{Brauninger:2009pe}
Carolin~B. Brauninger and Marco Cirelli.
\newblock {Anti-deuterons from heavy Dark Matter}.
\newblock {\em Phys. Lett. B}, 678:20--31, 2009.

\bibitem{amssens}
Francesca Giovacchini and V.~Choutko.
\newblock {Cosmic Rays Antideuteron Sensitivity for AMS-02 Experiment}.
\newblock In {\em {30th International Cosmic Ray Conference}}, volume~4, pages
  765--768, 2008.

\bibitem{ALADINO}
Oscar Adriani, Corrado Altomare, Giovanni Ambrosi, Philipp Azzarello, Felicia
  Carla~Tiziana Barbato, Roberto Battiston, Bertrand Baudouy, Benedikt
  Bergmann, Eugenio Berti, Bruna Bertucci, Mirko Boezio, Valter Bonvicini,
  Sergio Bottai, Petr Burian, Mario Buscemi, Franck Cadoux, Valerio Calvelli,
  Donatella Campana, Jorge Casaus, Andrea Contin, Raffaello D'Alessandro,
  Magnus Dam, Ivan De~Mitri, Francesco de~Palma, Laurent Derome, Valeria
  Di~Felice, Adriano Di~Giovanni, Federico Donnini, Matteo Duranti, Emanuele
  Fiandrini, Francesco~Maria Follega, Valerio Formato, Fabio Gargano, Francesca
  Giovacchini, Maura Graziani, Maria Ionica, Roberto Iuppa, Francesco Loparco,
  Jes{\'u}s Mar{\'\i}n, Samuele Mariotto, Giovanni Marsella, Gustavo
  Mart{\'\i}nez, Manel Mart{\'\i}nez, Matteo Martucci, Nicol{\`o} Masi,
  Mario~Nicola Mazziotta, Matteo Merg{\'e}, Nicola Mori, Riccardo Munini,
  Riccardo Musenich, Lorenzo Mussolin, Francesco Nozzoli, Alberto Oliva,
  Giuseppe Osteria, Lorenzo Pacini, Mercedes Paniccia, Paolo Papini, Mark
  Pearce, Chiara Perrina, Piergiorgio Picozza, Cecilia Pizzolotto, Stanislav
  Posp{\'\i}{\v s}il, Michele Pozzato, Lucio Quadrani, Ester Ricci, Javier
  Rico, Lucio Rossi, Enrico~Junior Schioppa, Davide Serini, Petr Smolyanskiy,
  Alessandro Sotgiu, Roberta Sparvoli, Antonio Surdo, Nicola Tomassetti,
  Valerio Vagelli, Miguel~{\'A}ngel Velasco, Xin Wu, and Paolo Zuccon.
\newblock Design of an antimatter large acceptance detector in orbit (aladino).
\newblock {\em Instruments}, 6(2), 2022.

\bibitem{pdg}
Particle~Data Group, P~A Zyla, R~M Barnett, J~Beringer, O~Dahl, D~A Dwyer, D~E
  Groom, C~J Lin, K~S Lugovsky, E~Pianori, D~J Robinson, C~G Wohl, W~M Yao,
  K~Agashe, G~Aielli, B~C Allanach, C~Amsler, M~Antonelli, E~C Aschenauer, D~M
  Asner, H~Baer, Sw~Banerjee, L~Baudis, C~W Bauer, J~J Beatty, V~I Belousov,
  S~Bethke, A~Bettini, O~Biebel, K~M Black, E~Blucher, O~Buchmuller, V~Burkert,
  M~A Bychkov, R~N Cahn, M~Carena, A~Ceccucci, A~Cerri, D~Chakraborty, R~Sekhar
  Chivukula, G~Cowan, G~D'Ambrosio, T~Damour, D~de~Florian, A~de~Gouvêa,
  T~DeGrand, P~de~Jong, G~Dissertori, B~A Dobrescu, M~D'Onofrio, M~Doser,
  M~Drees, H~K Dreiner, P~Eerola, U~Egede, S~Eidelman, J~Ellis, J~Erler, V~V
  Ezhela, W~Fetscher, B~D Fields, B~Foster, A~Freitas, H~Gallagher, L~Garren,
  H~J Gerber, G~Gerbier, T~Gershon, Y~Gershtein, T~Gherghetta, A~A Godizov, M~C
  Gonzalez-Garcia, M~Goodman, C~Grab, A~V Gritsan, C~Grojean, M~Grünewald,
  A~Gurtu, T~Gutsche, H~E Haber, C~Hanhart, S~Hashimoto, Y~Hayato, A~Hebecker,
  S~Heinemeyer, B~Heltsley, J~J Hernández-Rey, K~Hikasa, J~Hisano, A~Höcker,
  J~Holder, A~Holtkamp, J~Huston, T~Hyodo, K~F Johnson, M~Kado, M~Karliner, U~F
  Katz, M~Kenzie, V~A Khoze, S~R Klein, E~Klempt, R~V Kowalewski, F~Krauss,
  M~Kreps, B~Krusche, Y~Kwon, O~Lahav, J~Laiho, L~P Lellouch, J~Lesgourgues,
  A~R Liddle, Z~Ligeti, C~Lippmann, T~M Liss, L~Littenberg, C~Lourengo, S~B
  Lugovsky, A~Lusiani, Y~Makida, F~Maltoni, T~Mannel, A~V Manohar, W~J
  Marciano, A~Masoni, J~Matthews, U~G Meißner, M~Mikhasenko, D~J Miller,
  D~Milstead, R~E Mitchell, K~Mönig, P~Molaro, F~Moortgat, M~Moskovic,
  K~Nakamura, M~Narain, P~Nason, S~Navas, M~Neubert, P~Nevski, Y~Nir, K~A
  Olive, C~Patrignani, J~A Peacock, S~T Petcov, V~A Petrov, A~Pich, A~Piepke,
  A~Pomarol, S~Profumo, A~Quadt, K~Rabbertz, J~Rademacker, G~Raffelt, H~Ramani,
  M~Ramsey-Musolf, B~N Ratcliff, P~Richardson, A~Ringwald, S~Roesler, S~Rolli,
  A~Romaniouk, L~J Rosenberg, J~L Rosner, G~Rybka, M~Ryskin, R~A Ryutin,
  Y~Sakai, G~P Salam, S~Sarkar, F~Sauli, O~Schneider, K~Scholberg, A~J
  Schwartz, J~Schwiening, D~Scott, V~Sharma, S~R Sharpe, T~Shutt, M~Silari,
  T~Sjöstrand, P~Skands, T~Skwarnicki, G~F Smoot, A~Soffer, M~S Sozzi,
  S~Spanier, C~Spiering, A~Stahl, S~L Stone, Y~Sumino, T~Sumiyoshi, M~J
  Syphers, F~Takahashi, M~Tanabashi, J~Tanaka, M~Taševský, K~Terashi,
  J~Terning, U~Thoma, R~S Thorne, L~Tiator, M~Titov, N~P Tkachenko, D~R Tovey,
  K~Trabelsi, P~Urquijo, G~Valencia, R~Van~de Water, N~Varelas, G~Venanzoni,
  L~Verde, M~G Vincter, P~Vogel, W~Vogelsang, A~Vogt, V~Vorobyev, S~P Wakely,
  W~Walkowiak, C~W Walter, D~Wands, M~O Wascko, D~H Weinberg, E~J Weinberg,
  M~White, L~R Wiencke, S~Willocq, C~L Woody, R~L Workman, M~Yokoyama,
  R~Yoshida, G~Zanderighi, G~P Zeller, O~V Zenin, R~Y Zhu, S~L Zhu,
  F~Zimmermann, J~Anderson, T~Basaglia, V~S Lugovsky, P~Schaffner, and W~Zheng.
\newblock {Review of Particle Physics}.
\newblock {\em Progress of Theoretical and Experimental Physics},
  2020(8):083C01, 08 2020.

\bibitem{AMBROSI2023168023}
G.~Ambrosi, M.~Ambrosio, C.~Aramo, B.~Bertucci, E.~Bissaldi, M.~Bitossi,
  A.~Boiano, C.~Bonavolontà, M.~Capasso, A.~Circiello, L.~Consiglio,
  D.~Depaoli, F.~{Di Pierro}, L.~{Di Venere}, E.~Fiandrini, N.~Giglietto,
  F.~Giordano, S.~Incardona, M.~Ionica, F.~Licciulli, S.~Loporchio,
  G.~Marsella, V.~Masone, F.R. Pantaleo, R.~Paoletti, B.~Ruggiero,
  A.~Rugliancich, P.~Silvestrini, L.~Stiaccini, J.~Tasseva, L.~Tosti,
  G.~Tripodo, V.~Vagelli, and M.~Valentino.
\newblock High-density near-ultraviolet silicon photomultipliers:
  Characterization of photosensors for cherenkov light detection.
\newblock {\em Nuclear Instruments and Methods in Physics Research Section A:
  Accelerators, Spectrometers, Detectors and Associated Equipment},
  1049:168023, 2023.

\bibitem{Coutu_2024}
S.~Coutu, P.S. Allison, M.~Baiocchi, J.J. Beatty, L.~Beaufore, D.H. Calderón,
  A.G. Castano, Y.~Chen, N.~Green, D.~Hanna, H.B. Jeon, S.B. Klein, B.~Kunkler,
  M.~Lang, R.~Mbarek, K.~McBride, S.I. Mognet, J.~Musser, S.~Nutter,
  S.~O'Brien, N.~Park, K.M. Powledge, K.~Sakai, M.~Tabata, G.~Tarlé, J.M.
  Tuttle, G.~Visser, S.P. Wakely, and M.~Yu.
\newblock The high energy light isotope experiment program of direct cosmic-ray
  studies.
\newblock {\em Journal of Instrumentation}, 19(01):C01025, jan 2024.

\bibitem{YUE201538}
Wang Yue, Chen Zongde, Li~Chenhui, He~Ran, Wang Shenyuan, Li~Baicheng, Wang
  Ruiheng, Liang Kun, Yang Ru, and Han Dejun.
\newblock Performance of ultra-small silicon photomultiplier array with active
  area of 0.12mm×0.12mm.
\newblock {\em Nuclear Instruments and Methods in Physics Research Section A:
  Accelerators, Spectrometers, Detectors and Associated Equipment}, 787:38--41,
  2015.
\newblock New Developments in Photodetection NDIP14.

\bibitem{dSiPM}
Inge Diehl, Finn Feindt, Karsten Hansen, Stephan Lachnit, Frauke Poblotzki,
  Daniil Rastorguev, Simon Spannagel, Tomas Vanat, and Gianpiero Vignola.
\newblock The desy digital silicon photomultiplier: Device characteristics and
  first test-beam results.
\newblock {\em Nuclear Instruments and Methods in Physics Research Section A:
  Accelerators, Spectrometers, Detectors and Associated Equipment},
  1064:169321, 2024.

\bibitem{EBAPS}
Jinzhou Bai, Bo~Wang, Yonglin Bai, Weiwei Cao, Yang Yang, Fanpu Lei, and Dan
  Su.
\newblock {Optimum design of electron bombarded active pixel sensor for
  low-level light single photon imaging}.
\newblock In Yadong Jiang, Xiaoliang Ma, Xiong Li, Mingbo Pu, Xue Feng, and
  Bernard Kippelen, editors, {\em 9th International Symposium on Advanced
  Optical Manufacturing and Testing Technologies: Optoelectronic Materials and
  Devices for Sensing and Imaging}, volume 10843, page 108430V. International
  Society for Optics and Photonics, SPIE, 2019.

\bibitem{Mazziotta_2025}
M.N. Mazziotta, A.R. Altamura, L.~Congedo, G.~De~Robertis, A.~Di~Mauro, J.O.
  Guerra-Pulido, F.~Licciulli, L.~Lorusso, P.~Martinengo, E.~Nappi,
  N.~Nicassio, G.~Paić, G.~Panzarini, R.~Pillera, and G.~Volpe.
\newblock Development of a novel compact and fast sipm-based rich detector for
  the future alice 3 pid system at lhc.
\newblock {\em Journal of Instrumentation}, 20(01):C01001, jan 2025.

\end{thebibliography}
\end{document}